\documentclass[preprint]{aastex7}
\usepackage{longtable}

\newcommand{\chandra}{{\it CHANDRA}}

\newcommand{\swift}{{\it Swift}}

\newcommand{\msun}{$M_{\odot}$}

\newcommand{\kps}{\mbox{km~s$^{-1}$}}

\begin{document}

\title{Revealing the accelerating wind in the inner region of colliding-wind binary WR112}

\author[orcid=0000-0002-3380-3307]{John D. Monnier}
\affiliation{Astronomy Department, University of Michigan, Ann Arbor, MI 48109, USA}
\email{monnier@umich.edu}  

% First block of authors
\author[orcid=0000-0002-2106-0403]{Yinuo Han}
\affiliation{Division of Geological and Planetary Sciences, California Institute of Technology, 1200 E. California Blvd., Pasadena, CA 91125, USA}
\email{yinuo@caltech.edu}

\author[orcid=0000-0002-7762-3172]{Michael F. Corcoran}
\affiliation{CRESST~II and X-ray Astrophysics Laboratory, NASA/Goddard Space Flight Center, Greenbelt, MD 20771, USA}
\affiliation{The Catholic University of America, 620 Michigan Ave., N.E. Washington, DC 20064, USA}
\email{michael.f.corcoran@nasa.gov}

\author[orcid=0000-0002-3601-6165]{Sanne Bloot}
\affiliation{ASTRON, Netherlands Institute for Radio Astronomy, Oude Hoogeveensedijk 4, Dwingeloo, 7991 PD, The Netherlands}
\affiliation{Kapteyn Astronomical Institute,University of Groningen, P.O. Box 800, 9700 AV, Groningen, The Netherlands}
\email{bloot@astron.nl}

\author[0000-0002-7167-1819]{Joseph R. Callingham}
\affiliation{ASTRON, Netherlands Institute for Radio Astronomy, Oude Hoogeveensedijk 4, Dwingeloo, 7991 PD, The Netherlands}
\affiliation{Anton Pannekoek Institute for Astronomy, University of Amsterdam, Science Park 904, 1098 XH Amsterdam, The Netherlands}
\email{callingham@astron.nl}

\author[]{William Danchi}
\affiliation{NASA Goddard Space Flight Center, Astrophysics Division, Greenbelt, MD, 20771, USA}
\email{william.c.danchi@nasa.gov}

\author[orcid=0000-0002-8186-4753]{Philip G. Edwards}
\affiliation{CSIRO Space and Astronomy, Australia Telescope National Facility, PO Box 76 Epping 1710, NSW, Australia}
\email{Philip.Edwards@csiro.au}

\author[]{Lincoln Greenhill}
\affiliation{Harvard-Smithsonian Center for Astrophysics, 60 Garden Street, Cambridge, MA 02138,}
\email{lgreenhill@cfa.harvard.edu}

\author[orcid=0000-0001-7515-2779]{Kenji Hamaguchi}
\affiliation{CRESST~II and X-ray Astrophysics Laboratory, NASA/Goddard Space Flight Center, Greenbelt, MD 20771, USA}
\affiliation{Department of Physics, University of Maryland, Baltimore County,1000 Hilltop Circle, Baltimore, MD 21250, USA}
\email{kenji.hamaguchi@gmail.com}

\author[0000-0001-9315-8437]{Matthew J. Hankins}
\affiliation{Arkansas Tech University, 215 West O Street, Russellville, AR 72801, USA}
\email{mhankins1@atu.edu}

\author[]{Ryan Lau}
\affiliation{NSF NOIRLab, 950 N. Cherry Ave., Tucson, AZ 85719, USA}
\email{ryan.lau@noirlab.edu}

\author[orcid=0000-0003-2869-7682]{Jon M. Miller}
\affiliation{Astronomy Department, University of Michigan, Ann Arbor, MI 48109, USA}
\email{jonmm@umich.edu}  

\author[]{Anthony F. J. Moffat}
\affiliation{D\'{e}partement de physique, Universit\'{e} de Montr\'{e}al, 1375 Avenue Th\'{e}r\`{e}se-Lavoie-Roux, Montr\'{e}al (QC), H2V 0B3, Qu\'{e}bec, Canada}
\email{anthony.f.j.moffat@umontreal.ca}

\author[]{Garreth Ruane}
\affiliation{Jet Propulsion Laboratory, California Institute of Technology, Pasadena, California, USA}
\email{garreth.ruane@jpl.nasa.gov}

\author[orcid=0000-0002-9213-0763]{Christopher M. P. Russell}
\affiliation{Department of Physics and Astronomy, Bartol Research Institute, University of Delaware, Newark, DE, 19716, USA}
\email{crussell@udel.edu}

\author[]{Anthony Soulain}
\affiliation{Univ. Grenoble Alpes, CNRS, Alpes, Grenoble, France}
\email{anthony.soulain@protonmail.com}

\author[orcid=0000-0002-1481-4676]{Samaporn Tinyanont}
\affiliation{National Astronomical Research Institute of Thailand, 260 Moo 4, Donkaew, Maerim, Chiang Mai, 50180, Thailand}
\email{samaporn@narit.or.th}

\author[orcid=0000-0001-7026-6291]{Peter Tuthill}
\affiliation{Sydney Institute for Astronomy
University of Sydney, NSW 2006,  Australia}
\email{peter.tuthill@sydney.edu.au}

\author[0000-0003-0774-6502]{Jason J. Wang}
\affiliation{Center for Interdisciplinary Exploration and Research in Astrophysics (CIERA) and Department of Physics and Astronomy, Northwestern University, Evanston, IL 60208, USA}
\email{jason.wang@northwestern.edu}

\author[0000-0002-8092-980X]{Peredur M. Williams}
\affiliation{Institute for Astronomy, University of Edinburgh, Royal Observatory, Edinburgh, EH9 3HJ, UK}
\email{pmw@roe.ac.uk}

%\collaboration{all}{The Terra Mater collaboration}

\begin{abstract}
Colliding winds in massive binaries generate X-ray–bright shocks, synchrotron radio emission, and sometimes even dusty ``pinwheel'' spirals.   We report the first X-ray detections of the dusty WC$+$O binary system WR 112 from Chandra and Swift, alongside 27 years of VLA/ATCA radio monitoring and new diffraction-limited Keck images. Because we view the nearly circular orbit almost edge-on, the colliding-wind zone alternates between heavy Wolf-Rayet wind self-absorption and near-transparent O-star wind foreground each 20-yr orbit, producing phase-locked radio and X-ray variability. This scenario leads to a prediction that the radio spectral index is flatter from a larger non-thermal contribution around the radio intensity maximum, which indeed was observed.
 Existing models that assume a single dust-expansion speed fail to reproduce the combined infrared geometry and radio light curve. Instead, we require an accelerating post-shock flow that climbs from near-stationary to $\sim$1350 km/s in about one orbital cycle, naturally matching the infrared spiral from 5'' down to within 0.''1, while also fitting the phase of the radio brightening. These kinematic constraints supply critical boundary conditions for future hydrodynamic simulations,  which can link hot-plasma cooling, non-thermal radio emission, X-ray spectra, and dust formation in a self-consistent framework. WR 112 thus joins WR~140, WR~104, and WR~70-16 (Apep) as a benchmark system for testing colliding-wind physics under an increasingly diverse range of orbital architectures and physical conditions.
\end{abstract}



\section{Introduction} 
Cosmic dust is essential for the formation of low-mass stars and habitable planets. We find dust forming in a variety of stellar environments:  explosive outbursts  (novae and  supernovae), slow-moving stellar outflows (red giant and supergiant winds), and in the fast winds of classical, hot, carbon-rich Wolf-Rayet (WR) stars (the evolved descendants of massive O-type stars).  In almost all of these outflows, dust production occurs under density, pressure and  chemical conditions that are poorly constrained.  

The production of dust in Wolf-Rayet systems is a particular mystery, since Wolf-Rayet stars have $T_{\rm{eff}}\sim50,000$~K and thus are strong sources of photo-ionizing, dust-destroying UV radiation -- yet somehow dust is able to form within this harsh environment.  Dust formation in WR stars seems to require not only a C-rich (WC-type) WR star, but quite likely also the presence of a binary companion in which the dust forms in the highly condensed colliding-wind zone between the two stars, although large clumping in turbulent winds might also trigger dust formation in some single WC stars \citep[e.g.,][]{daviduraz2012}.

Dust producing Wolf-Rayet binaries thus are especially important laboratories for studying the conditions needed for dust to condense, as we can find systems with different mass-loss rates and physical separations to explore the physics behind the dust formation process.  We often see patterns that repeat on the orbital period, allowing detailed study and follow-up. Most identified dust-producing WR binaries consist of a chemically evolved, carbon-rich Wolf-Rayet star, possessing a strong, dense radiatively-driven  stellar wind 
($\dot{M}\gtrsim10^{-6}$~\msun~yr$^{-1}$, $V_{\infty}\sim1000-3000$
\kps)
gravitationally bound to a massive (typically O-type) companion with a similarly fast (though less dense) radiatively-driven stellar wind.  The stellar winds from the two stars collide hypersonically, 
producing a strong, dense, extremely hot ($T>10^{7}$~K)  ``bow shock'', which emits in the thermal X-ray band (0.5~--~10~keV). 
The carbon-rich wind of the WR star is compressed in the colliding wind shock, and somehow the gas cools enough and remains dense enough that  PAH molecules in some cases \citep{Marchenko:1997kx}, carbon chains and amorphous carbon grains can form at a significant rate   \citep[$\sim10^{-10}-10^{-6}$~M$_\odot$ yr$^{-1}$,][]{lau2020}.   X-ray emission from the shock provides our only direct measure of the thermodynamic conditions (densities, temperatures, pressures and abundances) in the hot shocked gas,  the precursor to dust formation.  Non-thermal emission in the radio and X-ray regimes is also associated with this colliding wind zone \citep[e.g.,][]{moran1989,white1995, hamaguchi2018}.

Much of the above picture was worked out from infrared, radio and X-ray observations of WR~140, a highly-eccentric binary in a 8-yr orbit where dust is formed only near periastron passage \citep{williams1990}. Although the basic theory of colliding stellar winds was first understood in the 1990s \citep{usov1991,stevens1992}, we still struggle to fit the full WR~140 dataset, such as the time variable non-thermal radio emission \citep{white1995} and X-ray properties \citep{pollack2021} due to uncertainties in orbit, difficulty making unambiguous X-ray and non-thermal radio predictions from hydrodynamical models, and complications from the time-variable line-of-sight wind opacity in the dynamic circumstellar environment.

We can learn even more by combining radio and infrared photometry with diffraction-limited infrared imaging.  \citet{tuthill1999} made the unexpected discovery that the WC$+$O binary WR~104 looked like a spinning spiral on the sky, the first pinwheel nebula.  Later, WR~98a, WR~112 and others \citep{monnier1999,monnier2007} were added to this list, along with their radio properties \citep{monnier2002}.  Despite the clear evidence for colliding winds in this system, none of the {\em persistent} dust emitters had been detected in X-rays in surveys: The X-ray luminosity was lower than seen in more widely-separated colliding wind system WR140 (at apastron) and possibly more similar to the dust-free, close-in WR binary $\gamma^2$~Vel (P$\sim$79 days).  

The Wolf-Rayet binary WR112 is a particularly important example of a dust-producing long-period colliding wind binary.  While its period was first thought to be only a few years since it was a persistent dust-emitter, subsequent monitoring determined the period was much longer, around 20~years \citep{lau2020}.  Here we collect decades of unpublished radio and infrared monitoring along with the first X-ray detection to peer deeply into the inner regions of this colliding wind system.  In \S2, we present our current picture of WR~112 based largely on mid-infrared imaging.  In \S3, we present the first  X-ray detections of WR112 by the Chandra X-ray Observatory and the Neil Gehrels Swift Observatory, long-term radio monitoring from VLA and ATCA, multi-epoch diffraction-limited imaging with Keck aperture masking, and lastly a Keck adaptive optics image behind the vortex coronagraph.  In \S4, we extend the mid-infrared dust shell modeling to the inner region of the nebula ($<1$'') where we confront existing models with new data, finding severe discrepancies.  Here we compare our findings for WR112 with the prototypical colliding-wind binary WR~140. We then close with a brief set of conclusions in \S5.

\section{The WR~112 binary and colliding-wind model}
\label{wr112model}
\begin{figure}[htbp] %  figure placement: here, top, bottom, or page
   \centering
   \includegraphics[trim={6cm 15cm 6cm 0cm}, width=5.5 in]{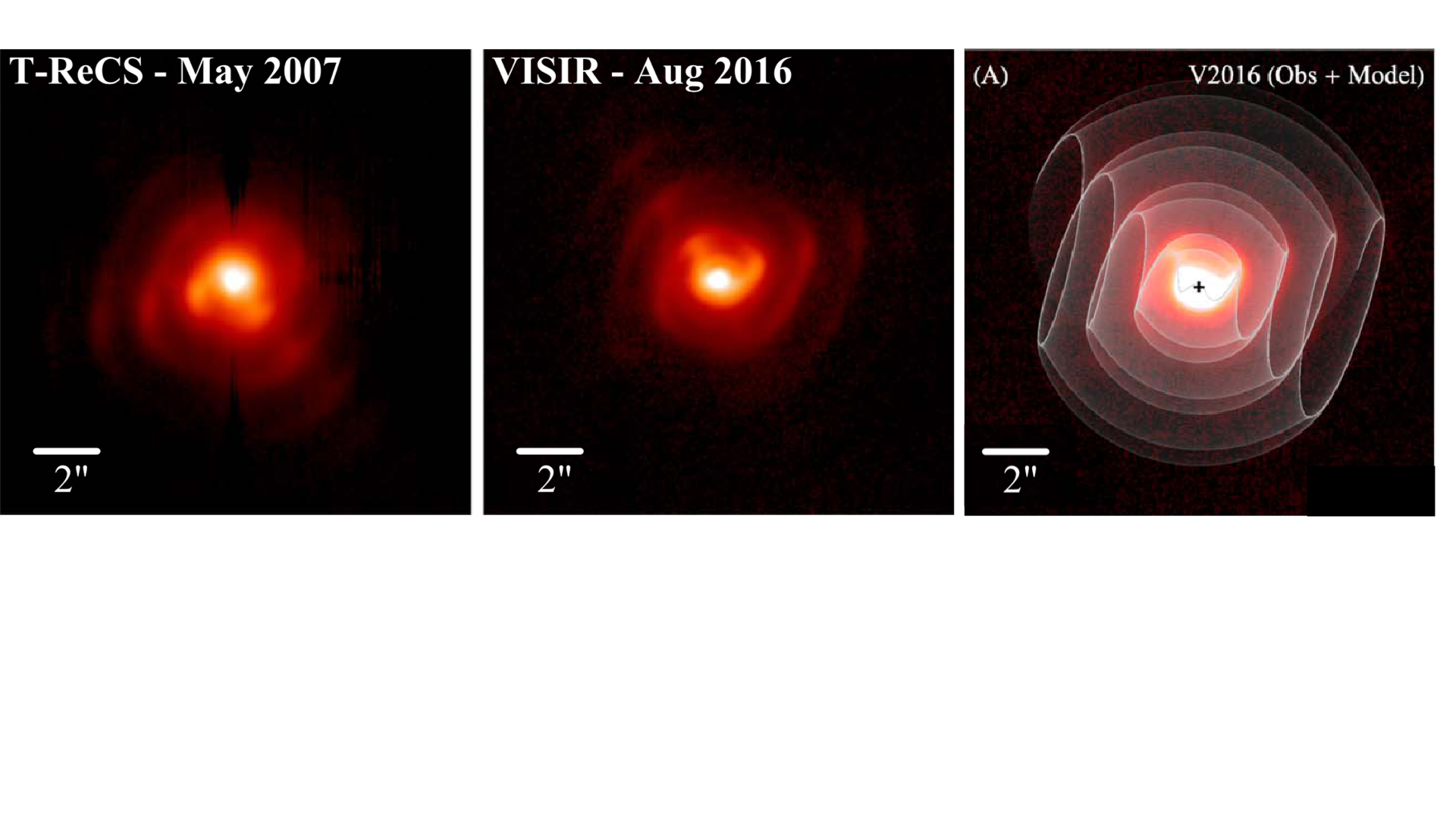}  
   \caption{\footnotesize{WR 112 N-band ($\lambda\sim12\mu$m) images from 2 epochs \citep[adapted from Figs. 2 \& 3 of][] {lau2020}. Note the different orientations of the U-shaped structures within $2''$ of the central binary, indicative of orbital motion. The rightmost figure shows a dust model applied to the 2016 N-band image. Orbital motion and the expansion of the dust arcs in the wind from WR 112 reproduces the observed spatial distribution of the inner and outer dust arcs projected on the sky.}}
   \label{fig:wr112n}
\end{figure}

WR 112 was first identified as a bright IR source from sounding-rocket observations made more than 40 years ago \citep{1978MNRAS.185...47C} and its dust shell was resolved by lunar occultation \citep{ragland1999}. This ``persistent'' dust emitter was first imaged in the mid-infrared by the Gemini telescope, revealing the presence of nested, extended, asymmetric dust arcs \citep{Marchenko:2002lr}.  These shells were erroneously interpreted as a face-on spiral, while radio and near-IR imaging would point to an edge-on geometry \citep{monnier2002, monnier2007}.   \citet{lau2020} settled the issue by detecting clear proper motions of the dust arcs from multiple mid-infrared images spanning most of the $P= 19.4\pm3$ yr orbital period. Although the binary itself has not been resolved yet, a 3-dimensional model of the dust was presented that suggests a nearly circular orbit with an inclination of  $i=100^{\circ}$ 
(see Figure~\ref{fig:wr112n}). 
Other long-period WC binaries have high orbital eccentricities and  tend to produce dust episodically only for some defined interval around periastron passage, while WR~112 continually produces dust \citep{williams1995} with constant infrared flux, suggesting a close-to-circular orbit.

\begin{table}
    \centering
    \caption{Geometrical parameters of the colliding wind model for WR~112.  %A more complex model is introduced in \S\ref{wind_modeling}.
    }
    \begin{tabular}{lcc}
        \hline
        & \citet{lau2020} & this work  \\\hline
        Binary Period $P$ [yr] & 19.4$^{+2.7}_{-2.1}$ & 20.0$\pm$0.1 \\
        Inclination $i$ [$^\circ$] & 100\,$\pm$\,15 & unchanged \\%\hline
        Longitude of asc. node $\Omega$ [$^\circ$] & -15\,$\pm$\,10 & unchanged \\
        Eccentricity $e$ & 0 & see \S\ref{wind_modeling}   \\
        Cone half-opening angle $\theta_w$ [$^\circ$] & 55\,$\pm$\,5 & unchanged \\
        Distance (kpc) & 3.39 & unchanged \\
        \hline
    \end{tabular}
    \label{table:parameters1}
\end{table}

The rate of dust production in WR 112 has been best measured as $M_{d} = $2.7$\times 10^{-6}$~\msun~yr$^{-1}$ by modeling of the dust arcs for an assumed distance of 3.4kpc \citep[][see Figure \ref{fig:wr112n}]{lau2020}. This is a much larger rate than the dust production rate of almost any other long-period WR binary previously studied \citep{lau2020}, and  comparable to that of WR104 %, the famous ``pinwheel binary'' 
\citep{tuthill1999}, a short period ($P=0.6 $~yr) system
which has a dust production rate of $4.4\times 10^{-6}$~\msun~yr$^{-1}$. Dust production for the eccentric WR140 binary system is strongly suppressed at wide separation, which makes it intriguing that WR~112 is such a prodigious dust producer.  The observed dust formation rate means that about 8\% of the wind of the WC star in WR 112 is converted to dust.  
The dust is consistent with amorphous carbon grains with a typical grain size $0.1-1.0\,\mu$m \citep{Marchenko:2002lr, lau2020}.

The
wind  of the WC star in WR 112 has a terminal velocity $V_{\infty, \mathrm{WC}}\sim1230\pm260$~\kps,  and a large wind-driven mass-loss rate, $\dot{M}_\mathrm{WC} \sim 1.1\times10^{-4}$~\msun~yr$^{-1}$ \citep{lau2020}, while the companion (assumed to be an early-O supergiant) probably has a higher wind velocity, $V_{\infty,\mathrm{O}}\sim2000$~\kps, with  a lower mass loss rate, $\dot{M}_\mathrm{O} \sim 8\times10^{-6}$~\msun~yr$^{-1}$.  At these velocities the thermalization temperature in the shock is $\sim$ 60~MK, with an expected luminosity of $\sim10^{33}$~ergs~s$^{-1}$ or so
(for comparison, the system has a bolometric luminosity $L_{bol}\approx$ 1.5
$\times10^{39}$~ergs~s$^{-1}$). 
Thus 
we need to also understand the thermodynamic conditions of the hot X-ray emitting gas in the colliding wind shock in order to fully understand how the colliding wind shock in WR 112 produces dust so efficiently. 
This requires measurement of the X-ray spectrum, since the X-ray spectrum probes the local conditions of the hot shocked gas near the point of impact,  free from contamination of other sources of emission associated with the stars or their winds.  

Table~\ref{table:parameters1} contains the overall geometry assumed for this paper based on modeling of the mid-infrared imaging.

\section{Observations}
\subsection{X-ray imaging and spectra}

\subsubsection{Neil Gehrels \swift\ Imaging Observations}

Unfortunately, few 
X-ray observations of 
WR~112 exist.  
X-ray emission from a source about 8.1$''$ from the known position of WR 112
was made 
in a short (853~s) exposure with the Neil Gehrels \swift\  X-ray Telescope (XRT) in photon-counting mode as part of a Director's Discretionary Time observation. 
This short exposure only detected 15
counts over the full XRT energy band (0.2~--10 keV, approximately). 
At the assumed distance of WR 112 (3.4~kpc) the XRT count rate corresponds to a substantial X-ray luminosity, $L_{x}\approx$ 1.8$\times 10^{33}$~ergs~s$^{-1}$. 
In 2022, the field centered on WR~112 was observed four additional times with the Swift XRT in photon-counting mode, confirming the detection of this source.
Table \ref{tab:xrayobs} lists these observations, along with the estimated observed X-ray luminosity at each epoch. 
The position of the \swift-XRT source was read off the \swift\ XRT image as 
$\alpha = 18h16m33.3s$, 
$\delta = -18^{\circ}$$58'$$34.7''$, which is about $8''$ from the 2mass J-band location of WR~112.  We also ran the \texttt{wavdetect} source detection tool (available through the Chandra CIAO analysis package) on the deepest Swift image, using initial wavelet scales of 2, 4, and 8 pixels.  This analysis returns a source position of $\alpha = 18h16m33.6s$, $\delta =-18^\circ58' 41.5''$, with a statistical error of about 0.6 arc~seconds.  This is formally $1.5''$ from the known location of WR 112, though systematic uncertainty in absolute alignment of the XRT image could not be quantified due to lack of other strong X-ray sources in the field.

\begin{deluxetable}{lcccc}
\tablecaption{X-ray Observations of WR 112}
\label{tab:xrayobs}
\tablehead{
\colhead{Facility ID} & \colhead{Start} & \colhead{End} & \colhead{Exposure} & \colhead{$L_X$}\\
\colhead{ } & \colhead{ } & \colhead{ } & \colhead{$\mathrm{s}$} & \colhead{10$^{33}$ erg s$^{-1}$}
}
\startdata
\swift\ 00014039002 & 2021-02-15T04:57 & 2021-02-15T05:11 & 849.08 & 1.8$\pm$0.6\\
\swift\ 00014039003 & 2022-05-04T20:12 & 2022-05-04T20:41 & 1730.61 & 2.9$\pm$0.5 \\
\swift\ 00014039004 & 2022-05-05T01:05 & 2022-05-05T02:59 & 1855.49 & 2.0$\pm$0.4 \\
\swift\ 00014039005 & 2022-05-08T15:21 & 2022-05-09T23:14 & 3024.25 & 2.5$\pm$0.3 \\
\swift\ 00014039006 & 2022-09-29T02:56 & 2022-09-29T23:52 & 5009.59 & 1.0$\pm$0.3 \\
\chandra\ 25129 & 2023-07-03T23:00 & 2023-07-04T04:37 & 17840.89 & 1.28$\pm$0.01\\
\chandra\ 27939 & 2023-07-04T10:47 & 2023-07-04T16:07 & 16748.81 & 1.18$\pm$0.04 \\
\enddata
\end{deluxetable}

\subsubsection{The \chandra\ Observations}\label{sec:obs}
In order to confirm that the \swift\ source was in fact WR 112, 
we 
observed the region around WR 112 on 2023 July 3~--~4 with \chandra. A list of the particular Chandra datasets can be accessed  in \dataset[DOI: X]{https://doi.org/10.25574/cdc.443}. We reprocessed the \chandra\ observations using the \texttt{chandra\_repro} python module distributed as part of the \texttt{ciao\_contrib.runtool} package, distributed with version 4.15.1 of the \texttt{ciao} analysis package.  Table \ref{tab:xrayobs} also summarizes the two \chandra\ pointings. The exposure time of the first observation segment, ObsID 25128 was 17.8~ks, while the exposure time of the second observation segment, ObsID 27939. was 16.7~ks. 

\begin{figure}[htbp] %  figure placement: here, top, bottom, or page
   \centering
   \includegraphics[width=6in]{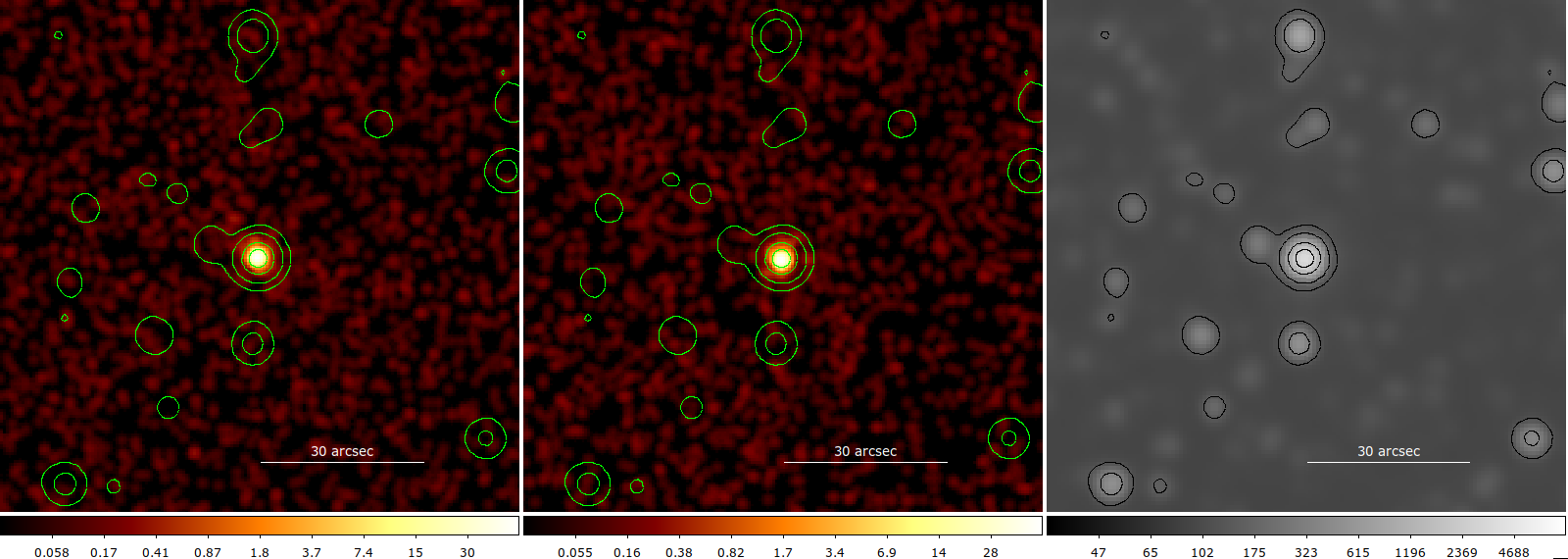} 
   \caption{Left: False-color Broad-band (0.2~--10~keV) \chandra\ image of the WR~112 field  from ObsID 25129; Center: \chandra\ image, ObsID 29739; Right: 2MASS J-band ($\lambda\sim1.23\mu$m) image of WR 112. Contours from the 2MASS image are overlaid on the \chandra images.  Note that, aside from WR 112, there are no other J-band sources clearly detected in the \chandra\ X-ray images.}
\label{fig:xopt}
\end{figure}

\begin{figure}[htbp] %  figure placement: here, top, bottom, or page
\centering
\includegraphics[width=5in]{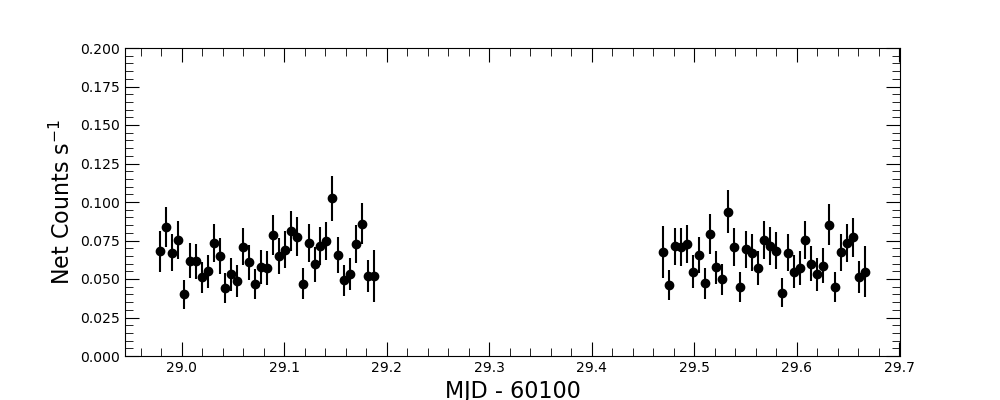} 
\caption{The 0.1~--~10~keV  band \chandra\ X-ray lightcurve of WR 112, using 500 second bins. \label{fig:xlc}
}
\end{figure}

Figure \ref{fig:xopt} compares a broad-band (0.2~--~10~keV) image of the WR 112 field from the \chandra\ observation ObsID \#25129, the 18~ksec observation, and the \chandra\ image from  ObsID 29739 with a J-band image from the 2MASS survey.  An alignment  correction of $0.5''$ was applied to ObsID \#29739 to align the strong X-ray source  in each X-ray image.  The brightest source in the 2MASS J-band image is WR~112.  Contours from the 2MASS J-band image are superimposed on the J-band image and on the two X-ray images.  The WR~112 J-band contour is well aligned with the strongest source in the \chandra\ X-ray images, confirming WR~112 as the X-ray source.  
Figure \ref{fig:xlc} shows the 0.1~--~10~keV band net count rates for WR 112 extracted from the \chandra\ observations. Source counts were extracted from a 12$''$ diameter circle centered on WR 112, and background was extracted from a circular source-free region of 51$''$ centered at $\alpha=18h:16m:38s$, $\delta=-18^{\circ}$:$58'$:$14''$. No obvious short-timescale variability is seen.  

We extracted spectra from Obsids 25129 and 27939 from circular regions of $12''$ centered on the optical position of WR~112.   %We extracted background from a $51''$ circular region centered at $\alpha = $ 18h16m38s, $\delta = -18^{\circ}$$58'$$14''$, containing no obvious X-ray sources. 
Backgrounds were extracted from nearby source-free regions.
We created redistribution matrix files (rmfs) and ancillary response files (arfs) for each Obsid using the CIAO \citep{ciao} \texttt{specextract} tool.  The individual spectra were revealed to be fully consistent in simple fits, so we added the spectra using the CIAO tool \texttt{combine\_spectra}.  The combined spectrum was then binned using the ``optimal'' binning algorithm of \cite{kaastra2016}, and fit over the 1--8~keV band using XSPEC version 12.14.1 \citep{arnaud1996}.  The fits minimized a Cash statistic \citep{cash1979}.

\begin{figure}[htbp] %  figure placement: here, top, bottom, or page
   \centering
\includegraphics[width=3.2in]{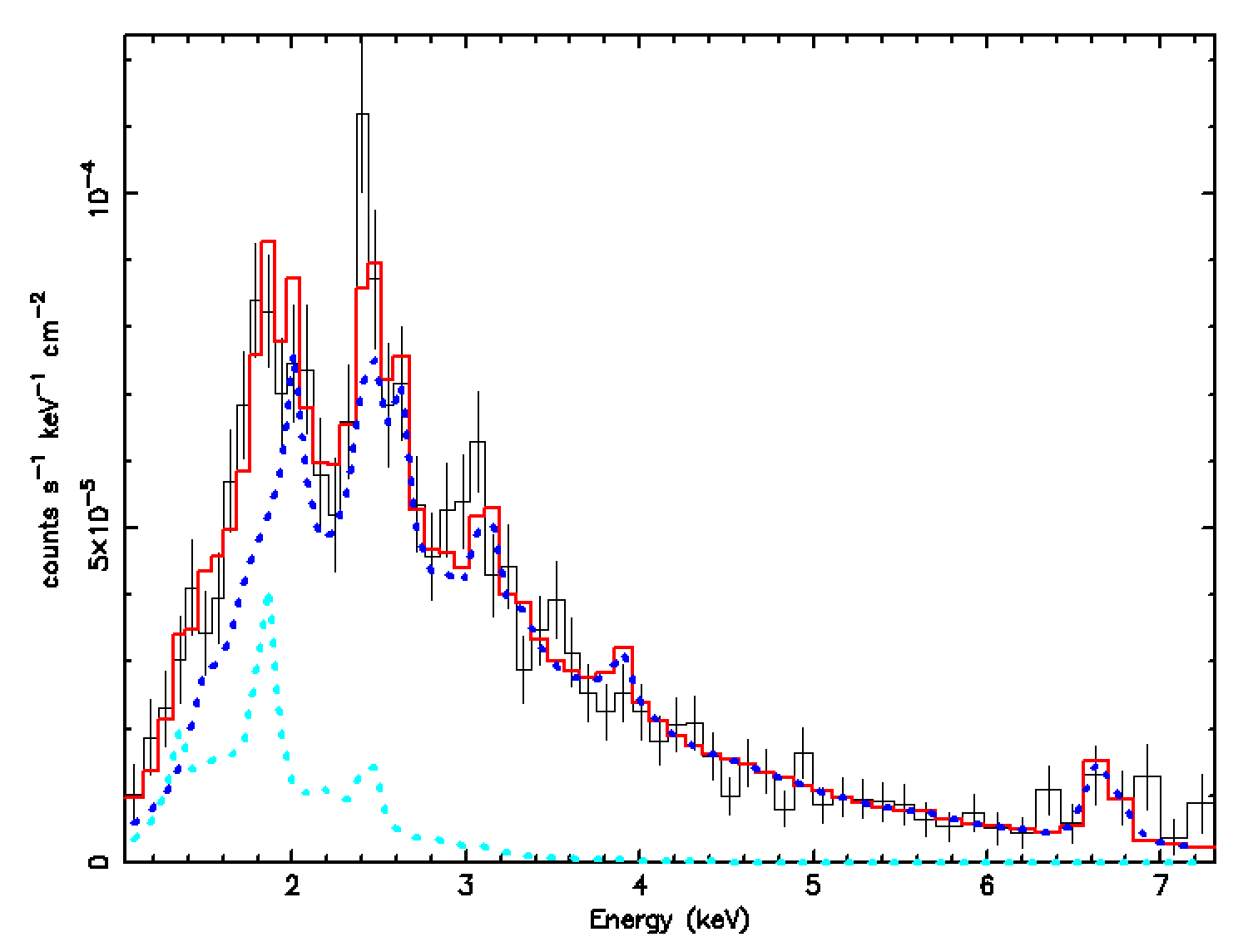} 
\caption{The summed Chandra ACIS X-ray spectrum of WR~112.  The data are binned for visual clarity.  The best-fit two-temperature plasma model is shown in red (\texttt{tbabs$\times$vapec$+$vapec}).  The kT$=0.5$~keV and kT$=2$~keV components are shown in cyan and blue, respectively.  The strongest lines in the model are He-like Si XIII, S XVI, Ar XVII, and Fe XXV (1.87, 2.46, 3.14, and 6.70~keV, respectively).  The spectrum is obscured at low energy by the high line-of-sight column density.  See Table 3 for model parameters and errors.}
   \label{fig:specx}
\end{figure}

We initially fit the combined spectrum using a simple single-temperature plasma model, modified by neutral gas absorption (\texttt{tbabs$\times$apec}).  This simple model provides an adequate fit ($C = 88$ for 59 degrees of freedom), but leaves residuals in the vicinity of He-like Si XIII (1.87~keV).  We therefore added a second plasma component and our spectral fit is shown in Figure~\ref{fig:specx}.  This yielded an improved fit ($C = 73$ for 57 degrees of freedom), and accounted for the potential Si~XIII flux.  This model over-predicts He-like Fe~XXV at 6.70~keV.  Replacing the simple plasma models with counterparts that have variable abundances (\texttt{tbabs$\times$vapec$+$vapec}) further improved the fit when the abundance of Fe is allowed to drift to sub-solar values ($C = 66$ for 56 degrees of freedom).  

The X-ray luminosity in the 1~--~10~keV band is comparable to the luminosity of  other WC-type colliding wind binaries, and is roughly equivalent to the X-ray luminosity of WR 140 when that binary is far from periastron passage \citep{pollack2021}.  The derived column density is also similar to the column to WR 140, when the star is far from periastron and absorption by the WC7 wind is unimportant.  Lastly, we note that the high temperature component (kT$\sim$2keV) is compatible with the  expected shock temperature of the WC wind given its terminal speed (1230 km/s).

\begin{deluxetable}{lc}
\tablecaption{X-ray Spectral Parameters Derived from the Chandra Spectrum Analysis}
\label{tab:spec}
\tablehead{\colhead{Parameter} & \colhead{Value} }
\startdata
$N_{\rm H}$ ($10^{22}$ cm$^{-2}$) & $3.0^{+0.5}_{-0.3}$ \\
\hline
kT (keV) & $0.5\pm 0.2$ \\
EM ($10^{56}$ cm$^{3}$) & $3^{+7}_{-1}$ \\
\hline
kT (keV) & $2.0\pm 0.2$ \\
EM ($10^{56}$ cm$^{3}$) & $2.4\pm 0.4$ \\
\hline
A$_{\rm Fe}$ & $0.4^{+0.2}_{-0.1}$ \\
\hline
unabsorbed $L_x$ ($10^{33}$ ergs s$^{-1}$, 1~--~10 keV) & $4.9\pm 0.5$ \\
\hline
Cash stat. & 65.7 \\
Deg. of freedom & 56 \\
\enddata
\end{deluxetable}

\subsection{Radio light curve}
Radio observations of WR~112 have been obtained from 1995 to 2022. These include 171 
flux density measurements 
not reported before by the principal investigators of those programs, and 15 previously published observations.  We note that some of the datasets presented here were analyzed from the Karl G. Jansky Very Large Array (VLA) archive and reported on by \cite{yam2015}, though our results here are independent and do not refer to this parallel effort.
 
In all observations, the radio emission was unresolved and here we report only flux measurements in a variety of bands, though mostly at X-band (3.6cm, 8GHz).
The detailed data analysis procedures were outlined in \citet{monnier2002} and followed throughout the observing campaign.  The older VLA data were analyzed using AIPS while the newer JVLA and ATCA data reduction employed CASA \citep{casa2022}. The main VLA flux amplitude calibrator was 3C286 while 1733-130, 1751-253, and 1832-105 were used for nearby phase calibrators.  The JVLA data taken after 2015 were calibrated with the standard VLA calibrator pipeline, with 3C286 as the amplitude calibrator and J1820-2528 as the phase calibrator. For ATCA observations, the amplitude calibrator was PKS~B1934-638 for the observations taken in 2020 and 2021. For the observation on 2022-03-06, PKS~B1934+638 was used as the amplitude calibrator for C- and X-band, while PKS~B1921-293 was used for the K-band data. PKS~B1730-130 was used as the phase calibrator.

\begin{figure}[htbp] %  figure placement: here, top, bottom, or page
   \centering
   \includegraphics[width=3.5in]{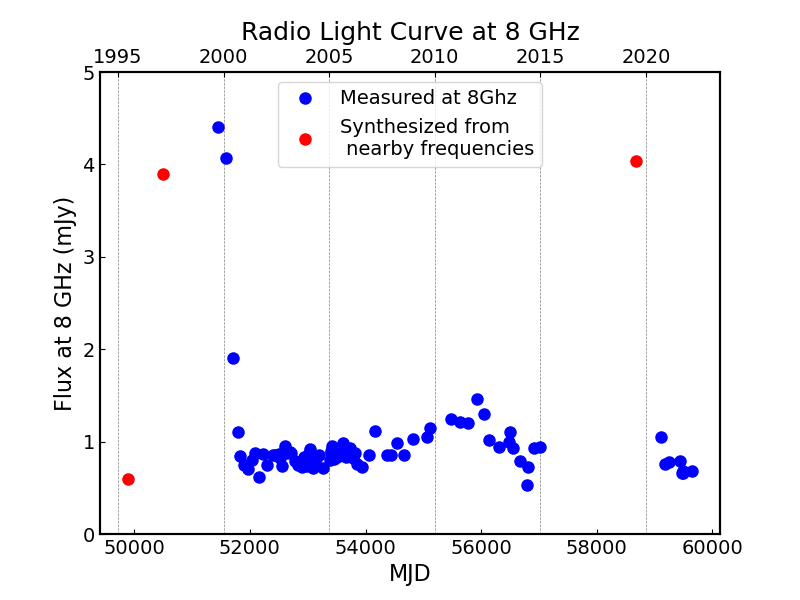} 
   \includegraphics[width=3.5in]{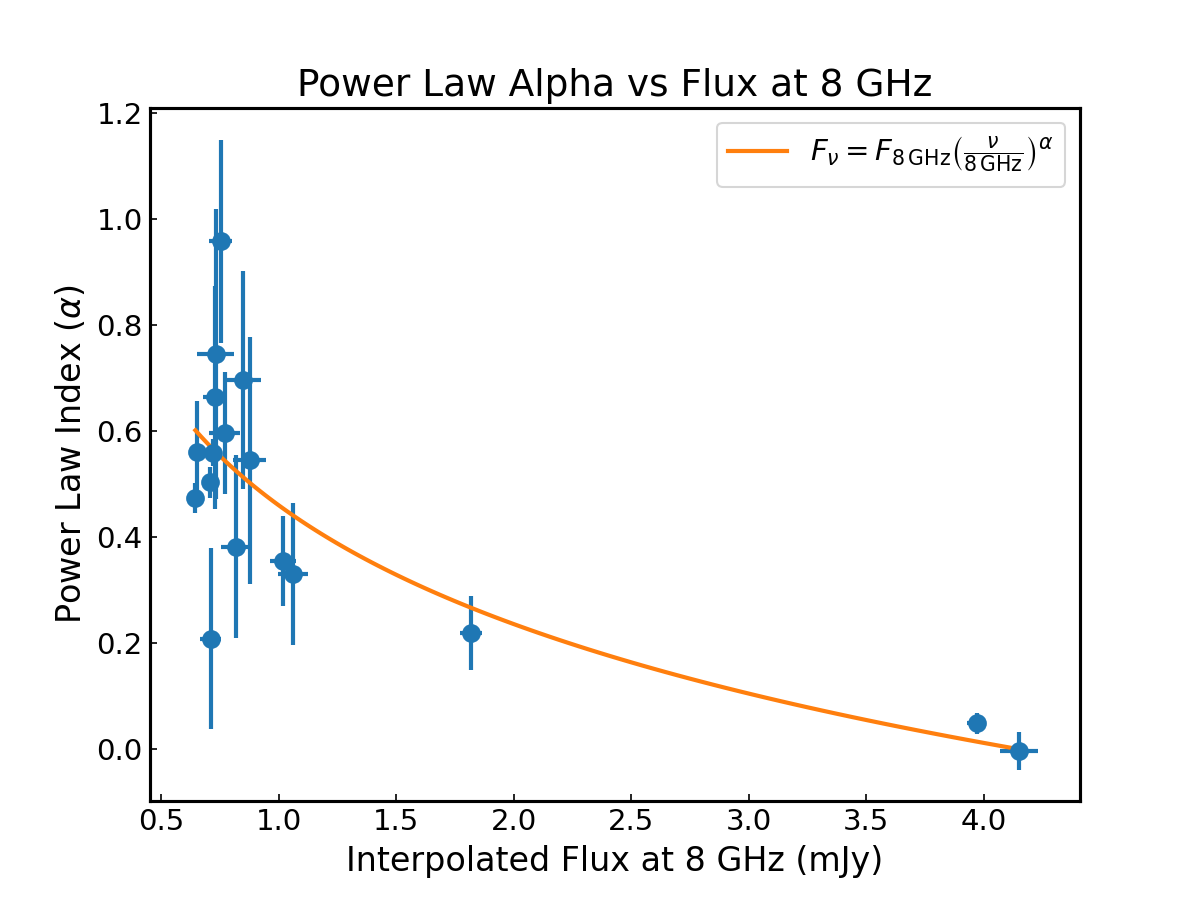}
   \caption{Left: The 8\,GHz light curve of WR112, including synthesized flux estimates when only nearby frequencies were available.  Right: The spectral index correlation with flux, as WR~112 transitions from the ``high state'' dominated by non-thermal radio emissions to the ``low state'' where we only see thermal (free-free) emission from the ionized winds.}
      \label{fig:radio1}
\end{figure}

Almost all the radio epochs include an 8GHz measurement, and the 8GHz flux curve 
is shown in Figure~\ref{fig:radio1} (left panel).  Note that the error bars are not included in the plot as they are typically smaller than the plot symbol.  When multiple wavelengths were available for a given epoch, we fit a simple power law model, $F_\nu = F_{\rm 8GHz} ( \frac{\nu}{\rm 8GHz})^\alpha$, between 2 and 25~GHz. Based on earlier work \citep[e.g.,][]{monnier2002}, we expect a thermal (free-free) spectrum ($\alpha\sim0.6$) when WR~112 is in the low state since the non-thermal radio emission from the colliding wind shock is buried behind the dense WR wind.  When the shock front is visible through the weaker O-star wind, we expect to see stronger radio flux with a flat spectrum ($\alpha\sim0$) characteristic of non-thermal synchrotron emission.  Figure~\ref{fig:radio1} (right panel) shows 
the  correlation between spectral index and flux density.  
Using this derived correlation, we synthesized 8~GHz flux estimates for observations missing measurements at 8~GHz. 
These points are included in the 8~GHz flux density curve shown in
%to 
Figure~\ref{fig:radio1} (left panel).  The full dataset with flux density measurements at all epochs (including error bars) and bands can be found in appendix \ref{appendix:radio}.

This long-term monitoring program began after \citet{monnier2002} highlighted the strong variability in WR~112 radio light curve based on the earlier data of \citet{leitherer1997} and \citet{chapman1999}.  
The flux declined from 4mJy down to $<$1mJy between 2000 February and 2000 October, accompanied by a dramatic change in the spectral index. 
Because the binary separation is too small to resolve directly, and the orbital velocities too low to determine a  radial velocity curve, constraining even the orbital period was difficult.  
Continued 
monitoring 
of the radio flux density 
offered the possibility of determining 
the orbital period of the 
binary, as had been done for WR~140 by \citet{white1995}.  Further, the percentage of the time the system is in the ``high state'' compared to the ``low-state'' 
can help constrain the inclination of the system and also the opening angle of the colliding wind shock. 
Based on earlier radio and dust observations of WR~140, we expected the period to be at most a few years.

We began a monitoring campaign which lasted for 15 years without return of the original radio  high state, and the VLA monitoring was terminated in 2015. The small rise in the radio emission in 2012--2013 was suspected as being due to a partial return to maximum light (not true) but it was unclear at the time.
 A renewed monitoring effort began in 2020 following an estimate of the orbital period of $19.4^{+2.7}_{-2.1}$~yrs from \cite{lau2020} based on the proper motion of mid-infrared dust shells.  
 
While 
our new monitoring began 
too late to catch the radio high state \citep[as confirmed by one 2019 measurement by the VLASS all-sky survey;][]{gordon2021}, 
the photometry shows 
that an observation 
on 2020 Sep 15, with an elevated flux above 1mJy, 
matches closely 
a similarly elevated flux observed on 
2000 Sep 08, almost 20 years earlier.
Assuming that the radio flux curve repeats itself exactly every orbit, 
these two observations 
suggest that the radio 
period is $20.0\pm0.1$~yrs, consistent with the period derived by \cite{lau2020} in the mid-IR. 
We note again the statistically-significant rise in radio brightness around 2012 that we can not currently explain, but might be understood with additional modelling.

\begin{figure}[htbp] %  figure placement: here, top, bottom, or page
   \centering
   \includegraphics[width=3.5in]{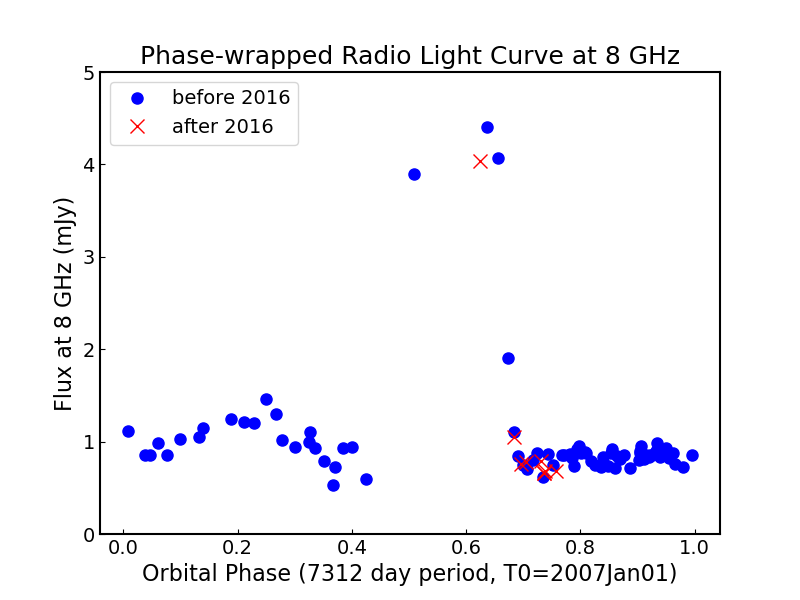} 
   \caption{The phased-up 8\,GHz light curve for WR112 (including synthesized flux estimates when only nearby frequencies were available) compared to orbital phase, assuming a period of 7312 days (20.0yrs).  The data before 2016 are marked with circles while data after marked with crosses to show the level of repeatability.}
   \label{fig:phased}

\end{figure}

The phased light curve is shown in Figure~\ref{fig:phased} with an arbitrary 
T$_0$ =  2007 Jan 01 (MJD 54101.0) and $P=20.0\pm0.1$ yrs.
The onset of the next high state should occur between 2035 and 2037 
and the high state should last 
$\sim$3~--~5yrs (15\%~--~23\% of the orbit). Note that the inclination of the system 
given in \S\ref{wr112model} 
is 100$\pm$15$\arcdeg$ (so that the line-of-sight is nearly in the orbital plane).  The cone half-opening angle is believed to be  
55$\arcdeg$ (see Table\,\ref{table:parameters1}),
which would naively suggest a $\sim$30\% duty-cycle when the line-of-sight traverses the weaker O-star wind into the non-thermal radio emitting region.  The slight discrepancy in the measured duty cycle 
is likely
due to a lower 
orbital
inclination which would reduce the fraction of the time the system is viewed in the high-state. 
Interestingly, we will show later in this paper that the phasing
of the radio light curve is far different than expected from the canonical wind model, and that will be key to deriving new model constraints on the wind acceleration to be discussed in \S\ref{wind_modeling}.

\subsection{Near-infrared imaging}
\subsubsection{Keck Aperture Masking}

WR~112 was observed eleven times between 1998 and 2004 at the Keck Observatory using the annulus aperture mask with the NIRC instrument \citep{matthews1994}. The first discoveries of ``pinwheel nebulae'' were made using the Keck masking experiment \citep{tuthill1999, monnier1999} and most of the Wolf-Rayet data has been published \citep{monnier2007}. 
 However, the full imaging of WR~112 has not appeared and here we present the full dataset for the first time.
 
Comprehensive explanations of the observing and imaging methodology for the Keck aperture masking experiment can be found in \citet{tuthill2000}, and we give only a short overview here.  An annulus aperture mask is attached in front of the Keck-1 secondary mirror and a series of $\sim$100 short exposure frames are collected.  The interference pattern can be analyzed using Fourier methods to extract the visibilities and closure phases for a range of baselines and closing triangles.  These observables are calibrated using contemporaneous observations of calibrators.  The pipeline produces OI-FITS data \citep{pauls2005} which can be used by a variety of image reconstruction packages, and here we used the MACIM algorithm \citep{ireland2006} for the images presented.

\begin{deluxetable}{llll}
\tablecaption{Keck Observatory Observing Log of WR112}
\label{kecktable}
\tablehead{\colhead{UT Date}  & \colhead{Instrument} & \colhead{Mode} & \colhead{Filter} }
\startdata
1998Jun05 &  NIRC & Aperture Masking: Annulus & ch4 ($\lambda_0~2.269\mu m,\Delta\lambda~0.155\mu m$) \\ 
1998Sep30 &  NIRC & Aperture Masking: Annulus & K ($\lambda_0~2.2135\mu m,\Delta\lambda~0.427\mu m$) \\ 
1999Apr26 &  NIRC & Aperture Masking: Annulus & K \\
1999Jul30 &  NIRC & Aperture Masking: Annulus & K \\
2000Jun24 &  NIRC & Aperture Masking: Annulus & K \\
2001Jun24 &  NIRC & Aperture Masking: Annulus & K \\ 
2002Jul24 &  NIRC & Aperture Masking: Annulus & K \\
2003May13 &  NIRC & Aperture Masking: Annulus & K \\
2004May29 &  NIRC & Aperture Masking: Annulus & H ($\lambda_0~1.6575\mu m,\Delta\lambda~0.333\mu m$) \\ 
2004May29 &  NIRC & Aperture Masking: Annulus & K \\
2004Sep04 &  NIRC & Aperture Masking: Annulus & K \\
\hline
2020Aug18 & NIRC2 & Vortex Coronagraph & Lp ($\lambda_0~3.776\mu m,\Delta\lambda~0.700\mu m$) \\ 
\enddata
\end{deluxetable}

\begin{figure}[htbp] %  figure placement: here, top, bottom, or page
   \centering
   \includegraphics[width=6in]{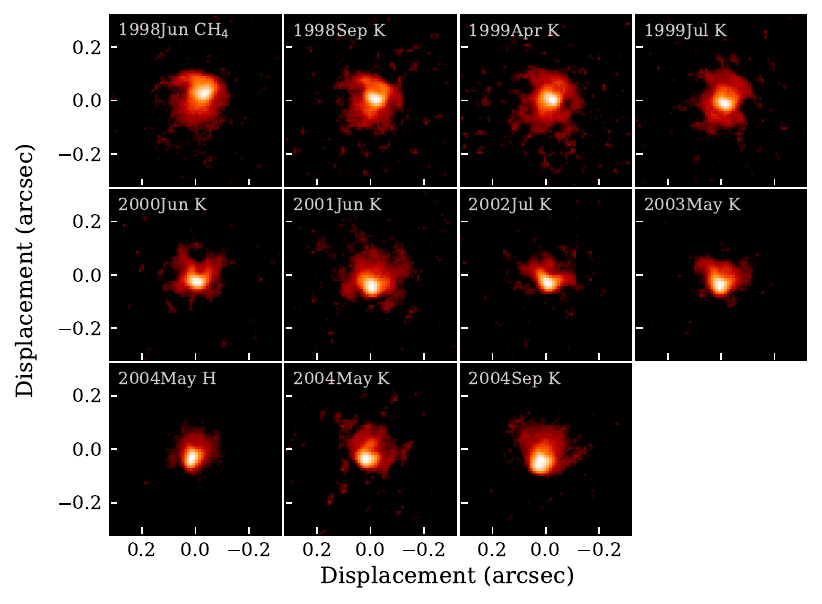} 
   \caption{Here we see near-infrared images of the inner arcsecond of WR112 created using aperture masking interferometry on the Keck Telescope. The asymmetry near the center reveals the direction the dust cone is facing on each date and allows to make a connection between the outer dust shells observed in the mid-infrared and inner ones (see modeling in \S\ref{wind_modeling}). 
   }\label{fig:keckmasking}
\end{figure}

Table\,\ref{kecktable}  contains the observing dates of the Keck aperture masking observations along with the observing filter used on each date. Note that most of the images were taken with an effective wavelength of $\sim$2.2$\mu$m.  Figure~\ref{fig:keckmasking} shows all the images in time order, revealing changes in the nebulosity and inner asymmetry over 6 years.  Qualitatively, these images will constrain inner geometry  in \S\ref{wind_modeling} when we connect the locations of the outer dust shells seen in the infrared with inner structure, along with the radio light curves which also constrains the innermost dust cone orientation over the orbit.

\subsubsection{Keck adaptive optics imaging}
L’-band (3.426-4.126 µm) observations of WR 112 were carried out with Keck/NIRC2 on UT2020Aug18 using the vortex coronagraph \citep{vargas2016, serabyn2017}. The observations employed the infrared pyramid wavefront sensor \citep{bond2020}, which operates at H-band, to control the Keck adaptive optics (AO) system. This was chosen due to heavy reddening of the target which precludes use of the Shack-Hartmann sensor which operates at R-band. The quadrant analysis of coronagraphic images for the tip-tilt sensing \citep[QACITS;][]{huby2017}   algorithm was used to keep the star aligned behind the mask by measuring tip/tilt residuals in the NIRC2 coronagraphic images and adjusting the tip/tilt offsets between the pyramid wavefront sensor and NIRC2. In total we obtained 87 frames, each consisting of 200 coadds of  0.053\,s of the target with the central star being obscured by the vortex coronagraph. This results in a total integration time of 922 s on source.  During the same night, we also observed five reference stars (HD162885, HD180732, HD181681, HD194479, HD194450) using the same vortex coronagraph setup. The only difference was that exposures consisting of 100 coadds of 0.18\,s integrations were used. All of these images were compiled into the reference PSF library for the imaging analysis.
Table~\ref{kecktable} contains the observing log information.

\begin{figure}[htbp] %  figure placement: here, top, bottom, or page
   \centering
   \includegraphics[width=6in]{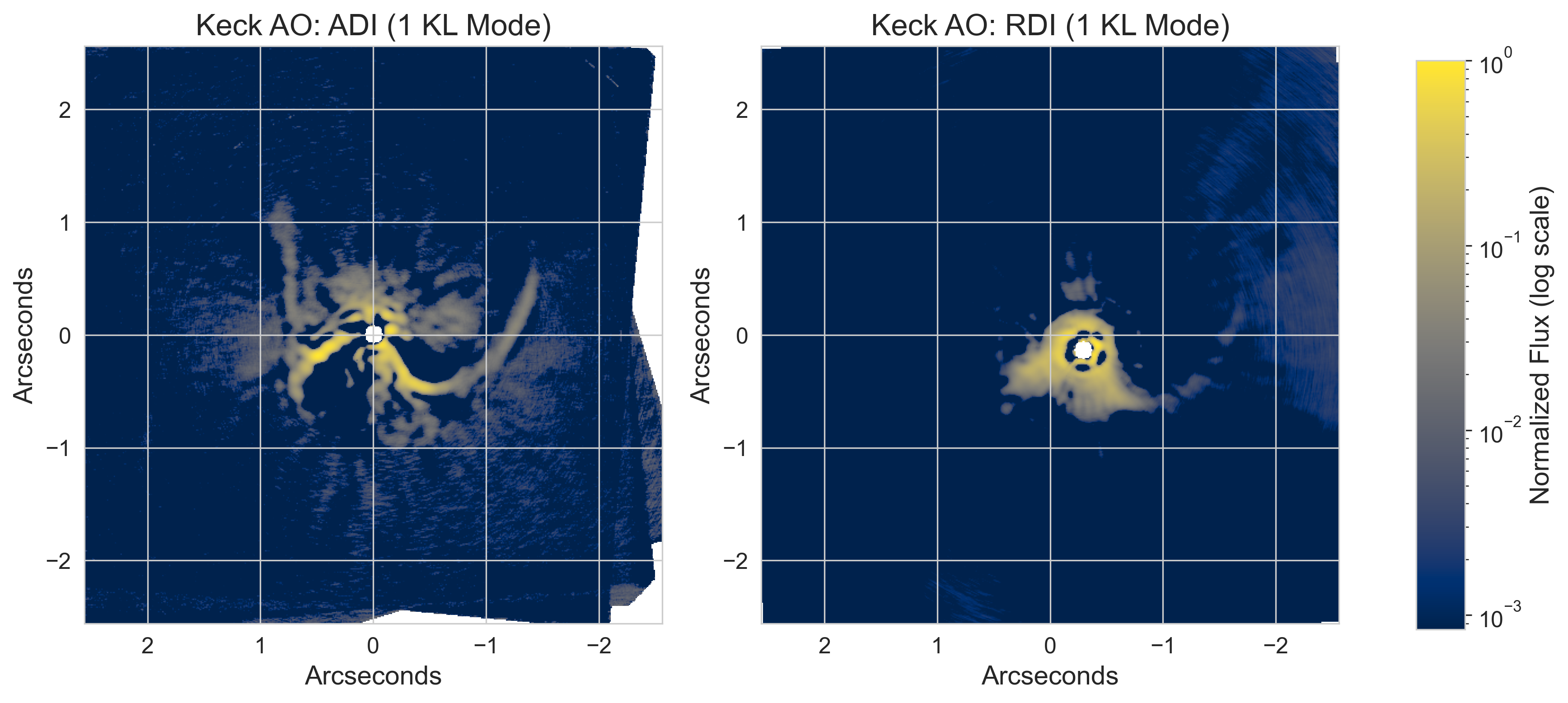} 
   \caption{Keck adaptive optics imaging with the vortex coronagraph shows the edge of the dust cone beautifully.  We present both the ADI and RDI images here and note that the central region is blocked by the coronagraph and is thus better revealed with Keck aperture masking images. The ADI reduction is more sensitive to sharp, faint features, while the RDI image better preserves the smooth, continuum features in the disk.  We model the orientation and shape of this dust cone in  \S\ref{wind_modeling}. }
   \label{fig:keckao}
\end{figure}

Initial raw data processing was done with the vortex pre-processing pipeline \citep{Xuan2018}. This pipeline performs bad pixel correction, flat field division, background subtraction, and relative image registration. The star was assumed to be centered, on average across the observations, behind the coronagraphic mask \citep{huby2017}. To remove the stellar point spread function (PSF), we tried both angular differential imaging (ADI; \citealt{Marois2006}) and reference star differential imaging (RDI). With ADI, images of WR112 taken at other times (i.e., other parallactic angles where the dust has rotated in the observed sky plane) are used to build up a model of the stellar PSF. ADI enables us to use PSFs taken close in time with very similar AO performance, but the disk itself leaks into the stellar PSF model and can subtract itself. WIth RDI, images of the five reference stars are used to model and subtract the stellar PSF. RDI avoids disk self-subtraction but the stellar PSF model is less accurate as well. In both cases, we use the KLIP implementation in the \texttt{pyKLIP} package for building up the model of the stellar diffraction pattern \citep{Wang2015}. For the ADI reductions, we built the KL modes across the entire frame, using any images of WR112 where the dust had rotated by at least 3 pixels due to ADI. For the RDI images, we also built the KL modes across the entire full frame, and used other vortex coronagraph images taken on the same night as reference images. We only use one KL mode to reconstruct the ADI and RDI images that appear in Figure~\ref{fig:keckao}.

The ADI image of WR112 shows the narrow limb brightened edge of the inner dust cone. This high resolution structure is very constraining for the wind models developed in \S\ref{wind_modeling}.  While ADI imaging can self-subtract nebulosity, we can see the presence of  close-in emission (to the south of central source) in the RDI image also presented in Figure\,\ref{fig:keckao}.

\section{Extending the Colliding-wind model}
\label{wind_modeling}
The diverse set of WR112 images is sensitive to dust structures over a range of radial regions from the binary. The near-IR aperture masking images ($\sim$0.1~arcsec), near-IR AO images ($\sim$1~arcsec) and mid-IR image ($\sim$5~arcsec) together offer valuable constraints on the binary orbit and dust dynamics within the system, which we aim to leverage by fitting a geometric model that consistently explains the available observations.

\subsection{Model with circular orbit and uniform speed}
%\subsubsection{Dust model}
The underpinnings of the colliding-wind dust model \citep{usov1991} was introduced in \S\ref{wr112model}. Here we remind readers of the basics before we add additional complications, including orbital eccentricity and a wind-acceleration zone.
We assume the column density distribution of the dust structure is produced on a cone-shaped wind-shock interface between the winds launched by the WR and O stars, after which the nucleated dust expands radially away from the binary along an imaginary conical surface. The orbital motion of the binary azimuthally offsets dust produced at one moment in time from the previous, resulting in a spiral dust surface in which all points expand radially away from the star.  In addition to the binary's orbital elements, the model is also specified by the half-opening angle of the cone and the expansion speed of the dust. Models based on the same mechanism have been applied to describe the structure of WR\,112 in the mid-IR \citep{lau2020} as well as in the context of other colliding-wind binaries such as WR140 \citep{han2022, lau2022},  and Apep \citep{callingham2019, han2020,bloot2022}. 

\citet{lau2020} fitted a dust surface model to multiple mid-infrared images of WR\,112, which revealed the motion of cool dust structures produced during approximately the three most recent orbital periods. We simulated the column density image of this model for the most recent orbital period of spiral dust structure. Compared to the mid-IR image, the near-IR AO image is sensitive to the high-resolution structure of much more close-in dust produced within the most recent orbital period. We find that the best-fit model parameters over-predict the size of this inner dust, but a delay in periastron passage by 0.3 orbits is able to simultaneously provide acceptable fits to both the AO and multi-epoch mid-IR images. We also simulated models of even closer-in dust structures at epochs that correspond to the aperture masking observations which are sensitive to the 0.1 arcsec scale, finding that this adjusted model reproduces the dust structures observed. We refer to this model with a circular orbit and uniform dust expansion speed as the ``original model'' and its parameters are displayed in Table~\ref{table:parameters2}. Note that the uncertainties were estimated by varying parameters independently and assessing the fit to the images by eye, rather than being quantified in a rigorous manner due to the lack of an effective metric \citep{han2020}.

\begin{table}
    \centering
    \caption{Parameters and plausibility of models invoked in this study. Additional parameters can be found in Table~\ref{table:parameters1}}
    \label{table:parameters2}
    \begin{tabular}{lcccc}
        \hline
        & Original & Eccentric & Accelerating & Ecc. + Acc. \\\hline
        Eccentricity $e$ & 0 & 0.4 & 0 & 0.2 \\
        Argument of periastron $\omega$ [$^\circ$] & 0 & 290\,$\pm$\,10 & 0 & 330\,$\pm$\,20 \\
        Epoch of periastron $T_0$ [Jul. yr] & 2021.0\,$\pm$\,1.9 & 2019.1\,$\pm$\,1.0 & 2012.3\,$\pm$\,1.0 & 2010.7\,$\pm$\,1.0 \\\hline
        Dust expansion speed $v$ [km\ s$^{-1}$] & 1220\,$\pm$\,60 & 1290\,$\pm$\,80 &  &  \\%\hline
        Initial velocity $v_0$ [km\ s$^{-1}$] & & & 0 & 0 \\
        Terminal velocity $v_\mathrm{d}$ [km\ s$^{-1}$] & & & 1360\,$\pm$\,100 & 1360\,$\pm$\,100 \\
        Acceleration phase [orbits] & & & 1.10\,$\pm$\,0.10 & 1.15\,$\pm$\,0.10 \\\hline
        Fits MIR geometry & Y & Y & Y & Y \\
        Fits NIR AO geometry & Y & Y & Y & Y \\
        Fits NIR masking geometry & Y & N & Y & Y \\
        Fits radio light curve & N & N & Y & Y \\\hline
    \end{tabular}
\end{table}

%\subsubsection{Wind model}
In addition to the imaging data available, the radio light curve also constrains the orbit of the binary. At an almost edge-on viewing perspective, the dominant stellar wind along the line of the sight, which absorbs free-free emission from the shock interface, alternates between that of the O and WR stars. The much lower optical depth of the O star's wind results in periodic brightening of the radio flux density, the timing of which is determined by the orbital configuration. Such a constraint likely probes even closer to the binary, since the free-free absorption coefficient is approximately proportional to wind density squared, and therefore $1/r^4$, implying that only the most close-in regions to the binary contribute to the optical depth.

To test whether the geometric model is consistent with the observed radio light curve, we modelled the column density of the O star's wind assuming that it occupies the volume within the shock cone. More specifically, rather than simulating rings of dust that form on and expand along conical surfaces, we simulated the stellar wind with concentric spherical caps, each consisting of uniformly distributed particles, and are azimuthally offset from those produced earlier in time due to orbital motion. The corresponding WR wind model is obtained by subtracting the O star wind model from a spherical wind model computed with the same $1/r^4$ scaling. 

Using such an approach, we find that the geometric model predicts the combined optical depth of the binary to dip, and therefore the radio flux to peak, from 2024 and 2028. However, the two epochs of radio observations imply that the radio flux should rise no earlier than 2015 and fall to quiescent levels no later than 2020. This implies that the geometric model differs from the observed radio peak by almost half an orbital period. 

Indeed, such a discrepancy might be expected based on an intuitive picture of the orbit of WR\,112. At mild eccentricities and for nearly edge-on orbits, the orbital phase at which the O star's stellar wind cone sweeps across the line of sight is primarily affected by $\omega$ and $T_0$ (and not by $\Omega$). Specifically, a near-transit (with the O star in front) of the two stars is achieved approximately a quarter of an orbit after $T_0$ for a circular orbit or with an offset based on the argument of periastron $\omega$ for an eccentric orbit. We therefore expect the radio peak to occur at approximately
\begin{equation}  \label{eq:radio}
    T_{\rm bright} \approx T_0 + P/4 - \frac{\omega}{360^\circ}P. 
\end{equation}
For this circular orbit model, we therefore expect the radio peak to be centred at approximately 2026, which the wind modelling confirms. In order for the radio peak to be consistent with the observations, we therefore require $T_0 - (\omega/360^\circ)P$ to be approximately between 2012 and 2013. A mechanism that alters the geometry of the spiral is therefore required to simultaneously explain all observations.

\subsection{Eccentric model}
Mild orbital eccentricity is able to warp the spiral geometry, as is understood to be the case in WR140 \citep{williams2009,han2022}, potentially allowing for a simultaneous fit to the imaging observations and radio flux. 
However, we experimented with a range of model parameters and found no combination that could explain all four of the AO, masking, mid-IR and radio constraints simultaneously. 

Eccentricity distorts the shape of the dust spiral in different ways at different points along the orbit, which implies that $\omega$ cannot be fixed at 0 as in the case of the original model. Simultaneously varying $\omega$ and $T_0$, we found that an eccentricity as high as 0.4 is still able to offer a close fit to the near-IR geometry while simultaneously reproducing the location of the mid-IR structures with the parameters listed in Table~\ref{table:parameters2}. However, such a model cannot reproduce the masking geometry, displaying a discrepancy much like a phase offset, reflecting that the warping introduced by eccentricity is inconsistent with the kind of warping suggested by the data. Furthermore, the predicted radio peak time from wind models that we correspondingly simulated predict radio brightening from 2024 to 2028, which is significantly different from the observations. An alternative mechanism to warp the basic spiral geometry is therefore required.

\subsection{Accelerating model}
\label{acceleration}
An expected process that is able to warp the close-in geometry and potentially explain these observations is a non-uniform expansion speed of dust. Acceleration of dust due to radiation pressure is expected to be important given the intense radiation field of the binary \citep[4$\times$10$^5$\,$L_\odot$;][]{sander2019}, and observational evidence for such an acceleration has been suggested in the context of WR140 \citep{han2022}. Assuming the same 19.4\,yr orbital period \citep{lau2020}, we modelled the variable expansion speed of dust to consist of two phases following its nucleation site assumed to be at the binary: a constant acceleration phase starting from a post-shock velocity of $v_0$ and reaching a velocity of $v_\mathrm{d}$ after an orbital phase of $\Delta \phi_a$, followed by expansion at a constant speed of $v_\mathrm{d}$.

\begin{figure}[htbp] %  figure placement: here, top, bottom, or page
   \centering
   \includegraphics[width=6in]{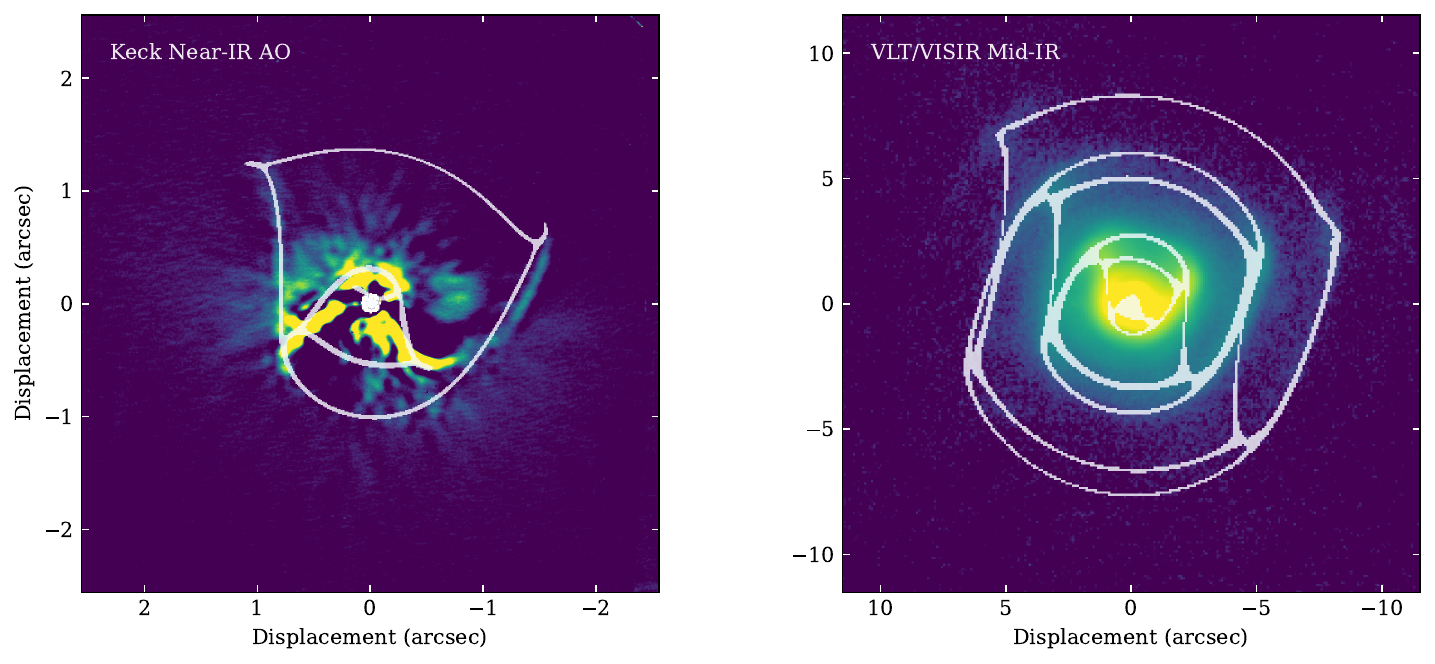} 
   \caption{Comparison of accelerating colliding-wind model prediction to the Keck near-infrared adaptive optics image and the VLT/VISIR mid-infrared image \citep{lau2022}. The model outline is shown with white lines, which are overlaid on the data. Visualizations of the model are provided in Figure~\ref{fig:visualisation}.} 
   \label{fig:imaging_comparison}
\end{figure}

\begin{figure}[htbp] %  figure placement: here, top, bottom, or page
   \centering
   \includegraphics[width=0.49\textwidth]{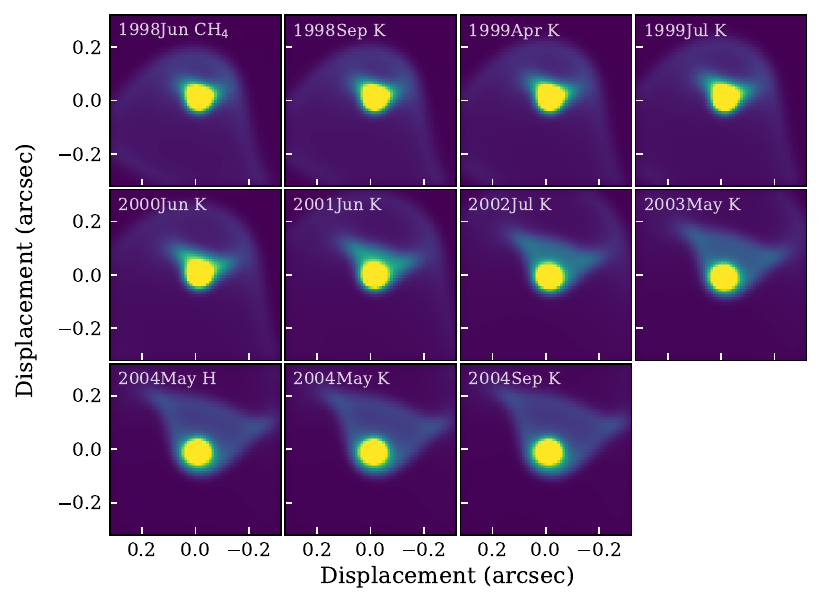} 
   \includegraphics[width=0.49\textwidth]{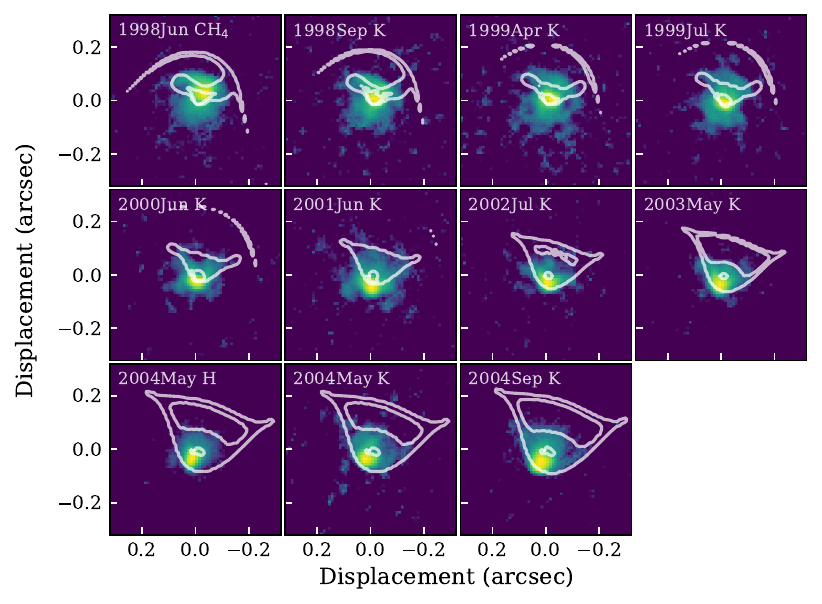} 
   \caption{Here we see model predictions (left panel) of the inner arcsecond dust emission from the accelerating wind model for each Keck masking dataset, along with original Keck masking images overlaid with white contours of the corresponding models (right panel). Models were convolved with K-band diffraction-limited point spread functions. 
   }
   \label{fig:masking_comparison}
\end{figure}

\begin{figure}[htbp] %  figure placement: here, top, bottom, or page
   \centering
   \includegraphics[width=3.5in]{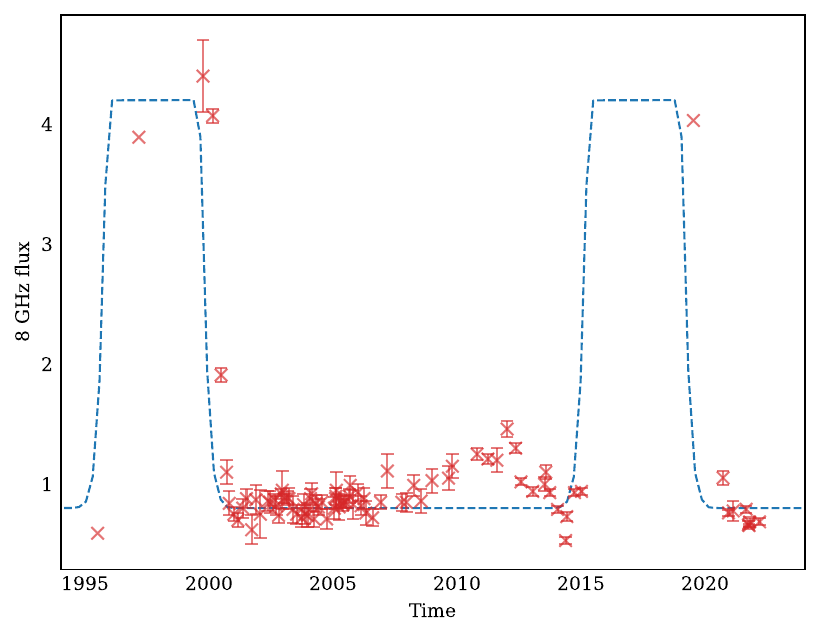} 
   \caption{The figure shows the observed radio flux measurements (red crosses) overlaid with the radio model (dashed blue line) that incorporates wind acceleration, as described in \S\ref{acceleration}. We roughly reproduce the start and stop times of the radio-bright phase while maintaining fits to the dust shell locations observed in the infrared. Note our simple model does not explain the mysterious rise and decline in flux around 2012.}
   \label{fig:radio_comparison}
\end{figure}

\begin{figure}
    \centering
    \includegraphics[width=1.0\textwidth]{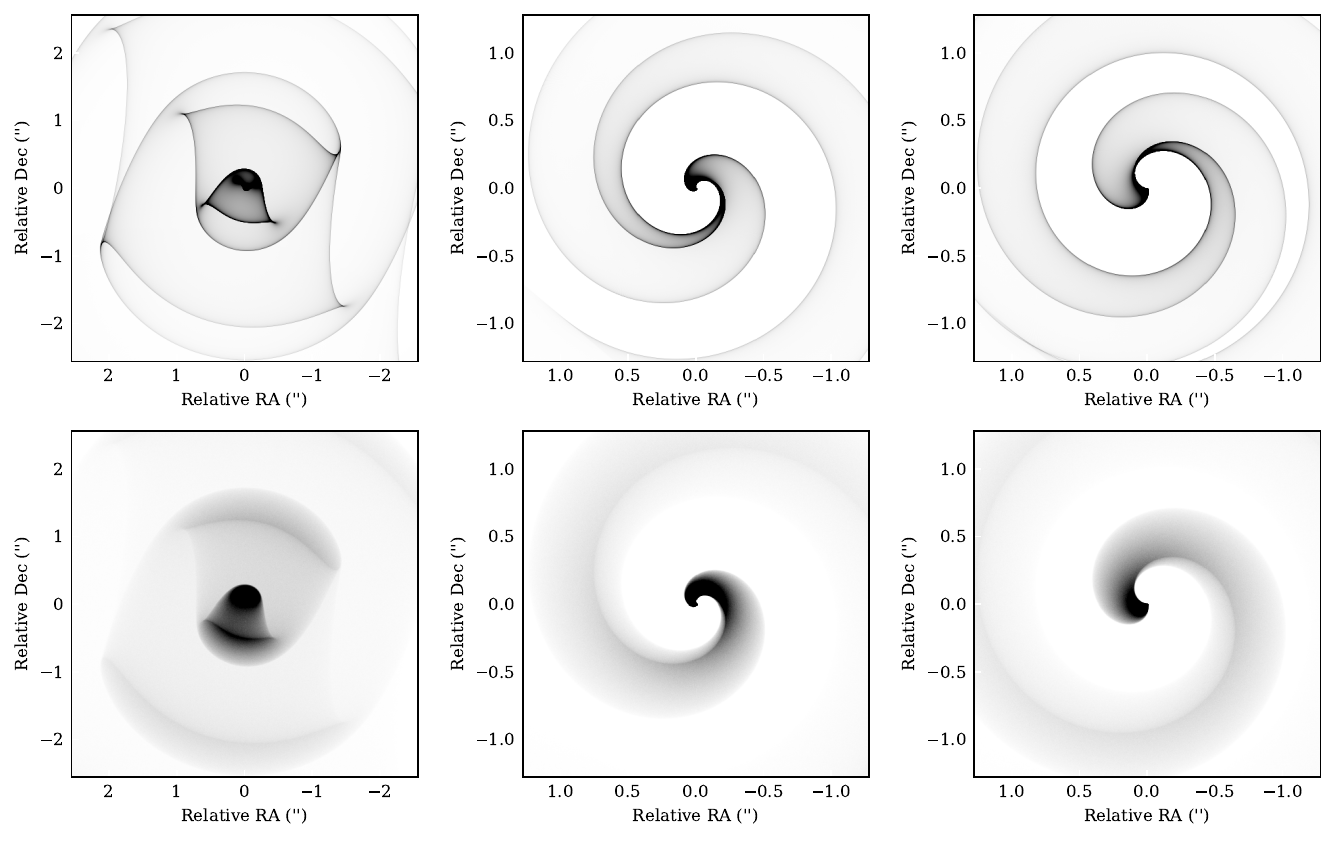}
    \caption{Visualization of the WR\,112 models simulated at the epoch of the near-IR AO observations. Top row: dust column density models. Bottom row: O star wind column density models. Left column: accelerating model of WR\,112 as viewed from Earth. Middle column: face-on view of the accelerating model (setting the observer at the left of the frame looking right). Right column: face-on models simulated without acceleration to visualize its effects by comparison with the middle column (observer at the left of the frame looking right) }
    \label{fig:visualisation}
\end{figure}

Motivated by Eq.~\ref{eq:radio}, we set the periastron passage time at 2012.3 to be consistent with expectations from the epoch of radio brightening while setting $e$ to be 0. With $i$, $\Omega$, $P$ and $\theta_w$ largely fixed by the multiple epochs of further out mid-IR geometry, the only free parameters are those defining the acceleration curve. We find that for the post-shock material that starts accelerating from a standstill ($v_0 = 0$), accelerating to 1360\ km\ s$^{-1}$ over approximately one orbit is able to simultaneously reproduce the mid-IR, AO and masking geometry. Terminal velocity is reached at a radius of approximately 3100\,au or 0.94$^{\prime\prime}$. Comparisons of the AO and mid-IR geometry between the accelerating model and the observations are shown in Figure~\ref{fig:imaging_comparison}. The masking model images along with a direct comparison to the Keck masking images are shown in Figure~\ref{fig:masking_comparison}. Furthermore, radio light curve modelling using our wind model does indeed produce brightening that is broadly consistent with the radio flux observations, as shown in Figure~\ref{fig:radio_comparison}. The best-fit model parameters are included in Table~\ref{table:parameters2}. We also present visualizations of the model for a few different viewing perspectives in Figure~\ref{fig:visualisation}.

Note that as the radio flux observations contain flux contributions (such as wind emission, which could contribute flux throughout the orbit) not accounted for in our model, a direct comparison of the absolute values predicted is unimportant for our purposes, which focuses on comparing the peak epoch of radio flux. 
We therefore calculated the radio light curve from the model as displayed in Figure~\ref{fig:radio_comparison}. We assumed that the radio flux is modulated by the optical depth of the free-free emission along the line of sight to the embedded non-thermal source, given by
\begin{equation}
    F_{\rm radio}(t) \propto \exp \left[ -( \tau_{\rm O}(t) + \tau_{\rm WR}(t) ) \right],
\end{equation}
where $\tau_{\rm O}$ and $\tau_{\rm WR}$ are the optical depth of the O and WR winds to free-free emission at the colliding-wind region, temporal variations of which are predicted by the wind column density model developed here. Note that the model only constrains $\tau_{\rm O}$ and $\tau_{\rm WR}$ in proportional terms and we do not constrain the wind densities directly here. Based on the radio spectral indices we are confident the WR wind is optically-thick and likewise that the O-star wind is optically-thin or not much above unity \citep[see deeper discussion by][]{monnier2002}.
We therefore focus on modeling the trends in the radio flux over time rather than its physical flux value, and $F_{\rm radio}(t)$ is linearly transformed to fit the flux values of the observations. This model is able to reproduce the switch in state between the radio maximum and minimum over an orbital period. Interestingly, a secondary brightening (2008 to 2013) is also observed but not predicted by the model, the nature of which will require further work to investigate given that its flux measurements presented here were obtained at only one frequency. 

While we assumed a circular orbit in this model, it is possible that the orbit of the binary is mildly eccentric. We experimented with the effect of incorporating a nonzero eccentricity into the dust and wind models, finding that a model largely consistent with all imaging and radio flux constraints could be achieved with the parameters under the ``eccentric + accelerating'' model in Table~\ref{table:parameters2}. 
While our accelerating models assume a post-shock velocity of $v_0 = 0$, dust is unlikely to form at a standstill, which is achieved only at the apex of the shock cone between the two stellar winds, but rather at tens to $\sim$100 au downstream \citep{tuthill2008, williams2009} following sufficient cooling. However, the phase of the binary, which is largely fixed by the radio light curve, implies that the geometry of the spiral as seen in the near-IR AO image (Figure~\ref{fig:keckao}) rapidly coils towards and disappears into the star when approaching the center. This geometry can only be reproduced with a $v_0$ that is significantly smaller than the terminal wind speed. We therefore set $v_0$ to 0 our model for simplicity. The conditions on the orbital phase imposed by the radio light curve can be partially relaxed with the introduction of a nonzero eccentricity, however we did not find any combination of orbital parameters that could simultaneously reproduce the imaging and radio light curve even if $v_0$ was set to only one third of the terminal wind speed. It is important to recognize that while the model aims to describe the dust geometry, what it fundamentally models is the dynamical evolution of material both before and after nucleating into dust. It is possible that this acceleration from a low initial velocity reflects the motion of material starting from post-shock wind near the apex of the shock cone before nucleating into dust. JWST observations \citep{richardson2025} may be able to offer sufficiently sharp resolution and high sensitivity to better constrain the eccentricity of the binary based on the geometry of further out dust shells in the mid-IR and its implications on the motion of post-shock wind and dust.

\subsection{Comparison with WR\,140}
The successful reproduction of available imaging and radio flux constraints in the accelerating model(s) for WR112 provides observational evidence of dust acceleration in a second WR binary after WR\,140. The two systems make for an interesting comparison. 

In the model for WR\,140 by \citet{han2022}, dust accelerates from approximately 1800 to 2500\ km\ s$^{-1}$ over approximately 2~yr, during which it travelled approximately 1,000\ au from the binary. Here in WR\,112, acceleration from approximately 0 to 1400\ km\ s$^{-1}$ occurs over 21~yr, or 6,000\ au from the binary. 
Since radiation pressure is proportional to stellar luminosity, which is approximately 4 times stronger in WR\,140 (1.43$\times$10$^6$\,$L_\odot$) compared to WR\,112 (4$\times$10$^5$\,$L_\odot$), the 5 times faster acceleration in WR\,140 compared to WR\,112 may therefore be largely consistent with differences in stellar luminosity. 

It is worth noting that key differences exist in the observations and modelling of the two systems. Whereas the acceleration parameters for WR\,112 from this study were primarily set by a single epoch of near-IR AO geometry in combination with orbital constraints from radio flux variations and the large-scale dust structure in the mid-IR, the orbit of WR\,140 is tightly constrained from relative astrometry and the acceleration was measured by cross-correlating features across a set of multi-epoch images which together track the expansion of each feature over time. Furthermore, the measured displacement-time diagram for WR\,140 was parameterized and fitted with a three-phase model, which included an earlier constant speed outflow phase before acceleration begins. In contrast, a model with reduced complexity was implemented for WR\,112 given the lack of a well-sampled displacement-time diagram shortly following dust production, which still offers the flexibility to model the shape of features in the near-IR AO image. Future work that better tracks the expansion of dust close to the binary will help provide more robust constraints on the dynamics of dust following their formation.

\section{Conclusions}
We have presented extensive new datasets for WR~112, including new X-ray, radio, aperture masking, and adaptive optics imaging.  This dust-producing colliding wind binary is in an unusually long orbit and viewed close to edge-on.  After a long period in a low-state, the radio flux recently increased then returned to quiescent levels, suggesting an orbital period of 20.0$\pm$0.1~yrs.  With multi-epoch imaging at different scales, we show that a simple uniform-outflow model can not fit the full dataset.  However, we do find that accelerating wind models fit the imaging and radio data although we do not constrain well the eccentricity nor the exact form of the acceleration flow in this work.

While  our geometrical models produce good qualitative fits to our diverse dataset, the parameters we present are not unique and still depend on a number of model assumptions. That said, our model is a major step forward in understanding WR~112 and will guide full hydrodynamical modeling and radiative transfer calculations to further constrain the physics of colliding wind binaries.  With a self-consistent physical model we can link the radio and X-ray properties to learn how high-temperature plasmas cool in these violent shocked flows between two massive stellar winds.  We add WR~112 to a small -- but growing -- list of colliding wind systems with sufficiently well-known properties suitable for detailed numerical modelling. Future work may wish to further explore post-shock kinematics and the cause of the secondary brightening of the radio light curve.

\facilities{Keck:I (NIRC), Keck:II (NIRC2), VLA, EVLA, ATCA, CXO, Swift}

\software{astropy \citep{2013A&A...558A..33A,2018AJ....156..123A,2022ApJ...935..167A}
          }

\appendix
\section{Radio Photometry}
\label{appendix:radio}
Here we include Table~\ref{tab:radio_obs} which contains the full radio photometry dataset used in this paper, including every known measurement in the literature.  When WR~112 was not detected, a 2-$\sigma$ upper limit is reported.

\startlongtable
\begin{deluxetable}{llllll}
\tablecaption{Radio Observations of WR112 (upper-limits are 2-$\sigma$)}
\label{tab:radio_obs}
\tablehead{\colhead{UT Date} & \colhead{Facility} & \colhead{$\lambda$ (cm)} & \colhead{$\nu$ (GHz)} & \colhead{$F_\nu$ (mJy)}  & \colhead{Reference} }
\startdata
1995Jun27 & ATCA & 3.0 & 10.0 & 0.68 $\pm$ 0.13 & \citet{leitherer1997} \\
1995Jun27 & ATCA & 6.0 & 5.0 & $<0.26$ & \citet{leitherer1997} \\
1997Feb25 & ATCA & 13.0 & 2.3 & 3.80 $\pm$ 0.16 & \citet{chapman1999} \\
1997Feb25 & ATCA & 20.0 & 1.5 & $<0.74$ & \citet{chapman1999} \\
1999Sep27 & VLA & 1.3 & 23.1 & 4.05 $\pm$ 0.25 & \citet{monnier2002} \\
1999Sep27 & VLA & 2.0 & 15.0 & 4.20 $\pm$ 0.30 & \citet{monnier2002} \\
1999Sep27 & VLA & 3.6 & 8.3 & 4.40 $\pm$ 0.30 & \citet{monnier2002} \\
1999Sep27 & VLA & 6.2 & 4.8 & 4.12 $\pm$ 0.10 & \citet{monnier2002} \\
1999Sep27 & VLA & 20.0 & 1.5 & 2.71 $\pm$ 0.17 & \citet{monnier2002} \\
2000Feb15 & VLA & 1.3 & 23.1 & 3.97 $\pm$ 0.12 & \citet{monnier2002} \\
2000Feb15 & VLA & 2.0 & 15.0 & 4.39 $\pm$ 0.17 & \citet{monnier2002} \\
2000Feb15 & VLA & 3.6 & 8.3 & 4.07 $\pm$ 0.06 & \citet{monnier2002} \\
2000Feb15 & VLA & 6.2 & 4.8 & 3.75 $\pm$ 0.08 & \citet{monnier2002} \\
2000Feb15 & VLA & 20.0 & 1.5 & 2.30 $\pm$ 0.30 & \citet{monnier2002} \\
2000Jun20 & VLA & 2.0 & 15.0 & 1.87 $\pm$ 0.16 & this work \\
2000Jun20 & VLA & 3.6 & 8.3 & 1.91 $\pm$ 0.06 & this work \\
2000Jun20 & VLA & 6.2 & 4.8 & 1.55 $\pm$ 0.08 & this work \\
2000Jun20 & VLA & 20.0 & 1.5 & 1.00 $\pm$ 0.20 & this work \\
2000Sep08 & VLA & 2.0 & 15.0 & 1.28 $\pm$ 0.13 & this work \\
2000Sep08 & VLA & 3.6 & 8.3 & 1.10 $\pm$ 0.10 & this work \\
2000Sep08 & VLA & 6.2 & 4.8 & 0.90 $\pm$ 0.08 & this work \\
2000Sep08 & VLA & 20.0 & 1.5 & $<0.90$ & this work \\
2000Oct17 & VLA & 2.0 & 15.0 & 1.02 $\pm$ 0.17 & this work \\
2000Oct17 & VLA & 3.6 & 8.3 & 0.84 $\pm$ 0.10 & this work \\
2000Oct17 & VLA & 6.2 & 4.8 & 0.68 $\pm$ 0.07 & this work \\
2000Oct17 & VLA & 20.0 & 1.5 & 0.55 $\pm$ 0.15 & this work \\
2000Dec23 & VLA & 0.7 & 42.9 & $<1.20$ & this work \\
2000Dec23 & VLA & 2.0 & 15.0 & 0.76 $\pm$ 0.11 & this work \\
2000Dec23 & VLA & 3.6 & 8.3 & 0.75 $\pm$ 0.06 & this work \\
2000Dec23 & VLA & 6.2 & 4.8 & 0.60 $\pm$ 0.10 & this work \\
2001Feb21 & VLA & 2.0 & 15.0 & 1.30 $\pm$ 0.20 & this work \\
2001Feb21 & VLA & 3.6 & 8.3 & 0.70 $\pm$ 0.06 & this work \\
2001Feb21 & VLA & 6.2 & 4.8 & 0.56 $\pm$ 0.06 & this work \\
2001Feb21 & VLA & 20.0 & 1.5 & 0.33 $\pm$ 0.10 & this work \\
2001May04 & VLA & 2.0 & 15.0 & 1.30 $\pm$ 0.30 & this work \\
2001May04 & VLA & 3.6 & 8.3 & 0.80 $\pm$ 0.08 & this work \\
2001May04 & VLA & 6.2 & 4.8 & 0.45 $\pm$ 0.05 & this work \\
2001May04 & VLA & 20.0 & 1.5 & 0.34 $\pm$ 0.11 & this work \\
2001Jun28 & VLA & 2.0 & 15.0 & 1.25 $\pm$ 0.18 & this work \\
2001Jun28 & VLA & 3.6 & 8.3 & 0.88 $\pm$ 0.08 & this work \\
2001Jun28 & VLA & 6.2 & 4.8 & 0.70 $\pm$ 0.15 & this work \\
2001Jun28 & VLA & 20.0 & 1.5 & 0.50 $\pm$ 0.18 & this work \\
2001Sep13 & VLA & 2.0 & 15.0 & 1.35 $\pm$ 0.25 & this work \\
2001Sep13 & VLA & 3.6 & 8.3 & 0.62 $\pm$ 0.12 & this work \\
2001Sep13 & VLA & 6.2 & 4.8 & 0.58 $\pm$ 0.10 & this work \\
2001Sep13 & VLA & 20.0 & 1.5 & $<0.82$ & this work \\
2001Nov13 & VLA & 2.0 & 15.0 & 1.35 $\pm$ 0.25 & this work \\
2001Nov13 & VLA & 3.6 & 8.3 & 0.87 $\pm$ 0.12 & this work \\
2001Nov13 & VLA & 6.2 & 4.8 & 0.61 $\pm$ 0.07 & this work \\
2001Nov13 & VLA & 20.0 & 1.5 & $<0.82$ & this work \\
2002Jan18 & VLA & 3.6 & 8.3 & 0.75 $\pm$ 0.20 & this work \\
2002May07 & VLA & 3.6 & 8.3 & 0.86 $\pm$ 0.08 & this work \\
2002Jun06 & VLA & 3.6 & 8.3 & 0.85 $\pm$ 0.09 & this work \\
2002Jun06 & VLA & 20.0 & 1.5 & $<0.28$ & this work \\
2002Aug17 & VLA & 3.6 & 8.3 & 0.87 $\pm$ 0.05 & this work \\
2002Aug17 & VLA & 20.0 & 1.5 & $<0.13$ & this work \\
2002Sep11 & VLA & 3.6 & 8.3 & 0.82 $\pm$ 0.08 & this work \\
2002Oct14 & VLA & 3.6 & 8.3 & 0.74 $\pm$ 0.06 & this work \\
2002Oct14 & VLA & 20.0 & 1.5 & $<0.39$ & this work \\
2002Nov22 & VLA & 3.6 & 8.3 & 0.92 $\pm$ 0.04 & this work \\
2002Nov22 & VLA & 20.0 & 1.5 & $<0.64$ & this work \\
2002Dec08 & VLA & 3.6 & 8.3 & 0.95 $\pm$ 0.16 & this work \\
2002Dec08 & VLA & 20.0 & 1.5 & $<0.50$ & this work \\
2003Jan05 & VLA & 3.6 & 8.3 & 0.88 $\pm$ 0.04 & this work \\
2003Feb22 & VLA & 3.6 & 8.3 & 0.89 $\pm$ 0.06 & this work \\
2003Mar17 & VLA & 3.6 & 8.3 & 0.88 $\pm$ 0.09 & this work \\
2003May15 & VLA & 2.0 & 15.0 & 1.90 $\pm$ 0.56 & this work \\
2003May15 & VLA & 3.6 & 8.3 & 0.79 $\pm$ 0.11 & this work \\
2003Jul06 & VLA & 3.6 & 8.3 & 0.75 $\pm$ 0.08 & this work \\
2003Jul06 & VLA & 20.0 & 1.5 & $<0.40$ & this work \\
2003Sep19 & VLA & 3.6 & 8.3 & 0.72 $\pm$ 0.08 & this work \\
2003Sep19 & VLA & 20.0 & 1.5 & $<0.33$ & this work \\
2003Oct17 & VLA & 3.6 & 8.3 & 0.83 $\pm$ 0.09 & this work \\
2003Oct17 & VLA & 20.0 & 1.5 & $<0.36$ & this work \\
2003Nov10 & VLA & 3.6 & 8.3 & 0.75 $\pm$ 0.08 & this work \\
2003Dec14 & VLA & 3.6 & 8.3 & 0.74 $\pm$ 0.10 & this work \\
2003Dec14 & VLA & 6.2 & 4.8 & 0.50 $\pm$ 0.10 & this work \\
2004Jan26 & VLA & 3.6 & 8.3 & 0.88 $\pm$ 0.08 & this work \\
2004Jan26 & VLA & 20.0 & 1.5 & $<0.62$ & this work \\
2004Feb13 & VLA & 3.6 & 8.3 & 0.92 $\pm$ 0.09 & this work \\
2004Mar20 & VLA & 3.6 & 8.3 & 0.71 $\pm$ 0.07 & this work \\
2004Apr16 & VLA & 3.6 & 8.3 & 0.84 $\pm$ 0.07 & this work \\
2004Apr16 & VLA & 20.0 & 1.5 & $<0.51$ & this work \\
2004May22 & VLA & 3.6 & 8.3 & 0.81 $\pm$ 0.07 & this work \\
2004May22 & VLA & 20.0 & 1.5 & $<0.35$ & this work \\
2004Jul12 & VLA & 3.6 & 8.3 & 0.85 $\pm$ 0.06 & this work \\
2004Jul12 & VLA & 20.0 & 1.5 & $<1.40$ & this work \\
2004Sep21 & VLA & 3.6 & 8.3 & 0.71 $\pm$ 0.08 & this work \\
2004Sep21 & VLA & 20.0 & 1.5 & $<0.26$ & this work \\
2005Jan12 & VLA & 3.6 & 8.3 & 0.80 $\pm$ 0.09 & this work \\
2005Jan12 & VLA & 20.0 & 1.5 & 0.51 $\pm$ 0.16 & this work \\
2005Jan30 & VLA & 3.6 & 8.3 & 0.88 $\pm$ 0.09 & this work \\
2005Jan30 & VLA & 20.0 & 1.5 & $<0.28$ & this work \\
2005Feb03 & VLA & 3.6 & 8.3 & 0.90 $\pm$ 0.07 & this work \\
2005Feb11 & VLA & 3.6 & 8.3 & 0.95 $\pm$ 0.15 & this work \\
2005Mar05 & VLA & 20.0 & 1.5 & 0.54 $\pm$ 0.23 & this work \\
2005Mar24 & VLA & 3.6 & 8.3 & 0.81 $\pm$ 0.11 & this work \\
2005Apr16 & VLA & 3.6 & 8.3 & 0.86 $\pm$ 0.05 & this work \\
2005May20 & VLA & 3.6 & 8.3 & 0.83 $\pm$ 0.05 & this work \\
2005Jun06 & VLA & 3.6 & 8.3 & 0.85 $\pm$ 0.05 & this work \\
2005Jul04 & VLA & 3.6 & 8.3 & 0.85 $\pm$ 0.05 & this work \\
2005Jul04 & VLA & 20.0 & 1.5 & $<0.32$ & this work \\
2005Aug22 & VLA & 3.6 & 8.3 & 0.90 $\pm$ 0.06 & this work \\
2005Aug22 & VLA & 20.0 & 1.5 & $<0.74$ & this work \\
2005Sep02 & VLA & 3.6 & 8.3 & 0.99 $\pm$ 0.08 & this work \\
2005Sep02 & VLA & 20.0 & 1.5 & $<0.70$ & this work \\
2005Oct16 & VLA & 3.6 & 8.3 & 0.83 $\pm$ 0.12 & this work \\
2005Dec24 & VLA & 3.6 & 8.3 & 0.93 $\pm$ 0.07 & this work \\
2005Dec24 & VLA & 20.0 & 1.5 & $<4.00$ & this work \\
2006Feb11 & VLA & 3.6 & 8.3 & 0.82 $\pm$ 0.08 & this work \\
2006Feb11 & VLA & 20.0 & 1.5 & $<0.26$ & this work \\
2006Mar25 & VLA & 3.6 & 8.3 & 0.88 $\pm$ 0.09 & this work \\
2006Mar25 & VLA & 20.0 & 1.5 & $<0.38$ & this work \\
2006Apr22 & VLA & 3.6 & 8.3 & 0.76 $\pm$ 0.10 & this work \\
2006Apr22 & VLA & 20.0 & 1.5 & $<0.58$ & this work \\
2006Jul30 & VLA & 3.6 & 8.3 & 0.72 $\pm$ 0.07 & this work \\
2006Jul30 & VLA & 20.0 & 1.5 & $<0.40$ & this work \\
2006Nov25 & VLA & 3.6 & 8.3 & 0.85 $\pm$ 0.06 & this work \\
2006Nov25 & VLA & 20.0 & 1.5 & $<1.20$ & this work \\
2007Mar05 & VLA & 3.6 & 8.3 & 1.11 $\pm$ 0.14 & this work \\
2007Oct03 & VLA & 3.6 & 8.3 & 0.85 $\pm$ 0.07 & this work \\
2007Dec09 & VLA & 3.6 & 8.3 & 0.85 $\pm$ 0.08 & this work \\
2008Mar26 & VLA & 3.6 & 8.3 & 0.99 $\pm$ 0.09 & this work \\
2008Jul12 & VLA & 3.6 & 8.3 & 0.86 $\pm$ 0.10 & this work \\
2008Dec21 & VLA & 3.6 & 8.3 & 1.03 $\pm$ 0.10 & this work \\
2009Aug21 & VLA & 3.6 & 8.3 & 1.05 $\pm$ 0.10 & this work \\
2009Oct17 & VLA & 3.6 & 8.3 & 1.15 $\pm$ 0.10 & this work \\
2010Oct15 & VLA & 3.6 & 8.3 & 1.25 $\pm$ 0.05 & this work \\
2011Mar22 & VLA & 3.6 & 8.3 & 1.21 $\pm$ 0.04 & this work \\
2011Aug03 & VLA & 3.6 & 8.3 & 1.20 $\pm$ 0.10 & this work \\
2012Jan02 & VLA & 3.6 & 8.3 & 1.46 $\pm$ 0.07 & this work \\
2012May04 & VLA & 3.6 & 8.3 & 1.30 $\pm$ 0.04 & this work \\
2012Jul23 & VLA & 3.6 & 8.3 & 1.02 $\pm$ 0.03 & this work \\
2013Jan12 & VLA & 3.6 & 8.3 & 0.94 $\pm$ 0.04 & this work \\
2013Jul05 & VLA & 3.6 & 8.3 & 1.00 $\pm$ 0.06 & this work \\
2013Jul23 & VLA & 3.6 & 8.3 & 1.10 $\pm$ 0.06 & this work \\
2013Sep20 & VLA & 3.6 & 8.3 & 0.93 $\pm$ 0.03 & this work \\
2014Jan13 & VLA & 3.6 & 8.3 & 0.79 $\pm$ 0.03 & this work \\
2014May11 & VLA & 3.6 & 8.3 & 0.53 $\pm$ 0.03 & this work \\
2014May29 & VLA & 3.6 & 8.3 & 0.73 $\pm$ 0.04 & this work \\
2014Sep18 & VLA & 3.6 & 8.3 & 0.93 $\pm$ 0.03 & this work \\
2014Dec31 & VLA & 3.6 & 8.3 & 0.94 $\pm$ 0.03 & this work \\
2019Jul06 & VLASS & 10.0 & 3.0 & 4.00 $\pm$ 0.50 & \citet{gordon2021} \\
2020Sep15 & ATCA & 1.4 & 21.2 & 1.45 $\pm$ 0.10 & this work \\
2020Sep15 & ATCA & 1.8 & 16.7 & 1.33 $\pm$ 0.08 & this work \\
2020Sep15 & ATCA & 3.3 & 9.0 & 1.05 $\pm$ 0.06 & this work \\
2020Sep15 & ATCA & 5.5 & 5.5 & 0.93 $\pm$ 0.12 & this work \\
2020Dec04 & VLA & 1.2 & 25.0 & 1.22 $\pm$ 0.06 & this work \\
2020Dec04 & VLA & 1.3 & 23.0 & 1.17 $\pm$ 0.07 & this work \\
2020Dec04 & VLA & 1.4 & 21.0 & 1.27 $\pm$ 0.06 & this work \\
2020Dec04 & VLA & 1.6 & 19.0 & 1.09 $\pm$ 0.05 & this work \\
2020Dec04 & VLA & 1.8 & 17.0 & 1.08 $\pm$ 0.07 & this work \\
2020Dec04 & VLA & 2.0 & 15.0 & 0.93 $\pm$ 0.03 & this work \\
2020Dec04 & VLA & 2.3 & 13.0 & 0.97 $\pm$ 0.03 & this work \\
2020Dec04 & VLA & 2.7 & 11.0 & 0.83 $\pm$ 0.03 & this work \\
2020Dec04 & VLA & 3.3 & 9.0 & 0.76 $\pm$ 0.03 & this work \\
2020Dec04 & VLA & 4.3 & 7.0 & 0.60 $\pm$ 0.03 & this work \\
2020Dec04 & VLA & 6.0 & 5.0 & 0.62 $\pm$ 0.03 & this work \\
2020Dec04 & VLA & 20.0 & 1.5 & 0.19 $\pm$ 0.05 & this work \\
2021Feb11 & ATCA & 1.4 & 21.2 & 1.35 $\pm$ 0.10 & this work \\
2021Feb11 & ATCA & 1.8 & 16.7 & 1.23 $\pm$ 0.08 & this work \\
2021Feb11 & ATCA & 3.3 & 9.0 & 0.78 $\pm$ 0.08 & this work \\
2021Feb11 & ATCA & 5.5 & 5.5 & 0.70 $\pm$ 0.15 & this work \\
2021Aug20 & VLA & 1.2 & 25.0 & 1.32 $\pm$ 0.05 & this work \\
2021Aug20 & VLA & 1.3 & 23.0 & 1.27 $\pm$ 0.05 & this work \\
2021Aug20 & VLA & 1.4 & 21.0 & 1.29 $\pm$ 0.05 & this work \\
2021Aug20 & VLA & 1.6 & 19.0 & 1.20 $\pm$ 0.03 & this work \\
2021Aug20 & VLA & 1.8 & 17.0 & 1.04 $\pm$ 0.04 & this work \\
2021Aug20 & VLA & 2.0 & 15.0 & 1.06 $\pm$ 0.03 & this work \\
2021Aug20 & VLA & 2.3 & 13.0 & 0.94 $\pm$ 0.03 & this work \\
2021Aug20 & VLA & 2.7 & 11.0 & 0.87 $\pm$ 0.03 & this work \\
2021Aug20 & VLA & 3.3 & 9.0 & 0.79 $\pm$ 0.03 & this work \\
2021Aug20 & VLA & 4.3 & 7.0 & 0.65 $\pm$ 0.03 & this work \\
2021Aug20 & VLA & 6.0 & 5.0 & 0.52 $\pm$ 0.04 & this work \\
2021Sep24 & VLA & 2.7 & 11.0 & 0.70 $\pm$ 0.03 & this work \\
2021Sep24 & VLA & 3.3 & 9.0 & 0.66 $\pm$ 0.03 & this work \\
2021Oct03 & ATCA & 1.4 & 21.2 & 0.98 $\pm$ 0.04 & this work \\
2021Oct03 & ATCA & 1.8 & 16.7 & 0.96 $\pm$ 0.02 & this work \\
2021Oct03 & ATCA & 3.3 & 9.0 & 0.66 $\pm$ 0.01 & this work \\
2021Oct03 & ATCA & 5.5 & 5.5 & 0.58 $\pm$ 0.02 & this work \\
2021Oct11 & VLA & 2.7 & 11.0 & 0.80 $\pm$ 0.04 & this work \\
2021Oct11 & VLA & 3.3 & 9.0 & 0.69 $\pm$ 0.04 & this work \\
2022Mar06 & ATCA & 1.4 & 21.2 & 0.94 $\pm$ 0.22 & this work \\
2022Mar06 & ATCA & 1.8 & 16.7 & 1.12 $\pm$ 0.12 & this work \\
2022Mar06 & ATCA & 3.3 & 9.0 & 0.69 $\pm$ 0.03 & this work \\
2022Mar06 & ATCA & 5.5 & 5.5 & 0.53 $\pm$ 0.03 & this work \\
\enddata
\end{deluxetable}

\begin{acknowledgments}
The authors would like to thank a few colleagues who have contributed to obtaining this data presented herein: Anthony Pollack,  Andreas Sander, and David Espinoza. We also thank the anonymous referee for their help in improving the manuscript.

Support for this work was provided by the National Aeronautics and Space Administration through Chandra Award Number 23200246 issued by the Chandra X-ray Center, which is operated by the Smithsonian Astrophysical Observatory for and on behalf of the National Aeronautics
Space Administration under contract NAS8-03060.

The National Radio Astronomy Observatory is a facility of the National Science Foundation operated under cooperative agreement by Associated Universities, Inc.  

The Australia Telescope Compact Array is part of the Australia Telescope National Facility (grid.421683.a) which is funded by the Australian Government for operation as a National Facility managed by CSIRO. We acknowledge the Gomeroi people as the traditional owners of the Observatory site.

Some of the data presented herein were obtained at Keck Observatory, which is a private 501(c)3 non-profit organization operated as a scientific partnership among the California Institute of Technology, the University of California, and the National Aeronautics and Space Administration. The Observatory was made possible by the generous financial support of the W. M. Keck Foundation.  
The authors wish to recognize and acknowledge the very significant cultural role and reverence that the summit of Maunakea has always had within the Native Hawaiian community. We are most fortunate to have the opportunity to conduct observations from this mountain. 

M.F.C. and K.H. are supported by NASA under award numbers 80GSFC21M0002 and  80GSFC24M0006. J.J.W. acknowledges support by the Heising-Simons Foundation (\#2023-4598) and the Alfred P. Sloan Foundation. S.B. acknowledges funding from the Dutch research council (NWO) under the talent programme (Vidi grant VI.Vidi.203.093). J.D.M. and M.C. acknowledge support from Chandra Award No. GO2-23012B. JRC acknowledges funding from the European Union via the European Research Council (ERC) grant Epaphus (project number: 101166008). CMPR acknowledges support from NASA Chandra Theory grant number TM3-24001X.

\end{acknowledgments}

\bibliography{wr112x,jdm_wr112}
\bibliographystyle{aasjournal}

\end{document}